\documentclass[10pt]{article}
\pdfoutput=1
\usepackage{graphicx,amssymb,amsfonts,amsmath,amssymb,amscd,amstext, mathrsfs,color}
\makeatletter \@addtoreset{equation}{section} \makeatother

\newcommand{\be}{\begin{eqnarray}}
\newcommand{\ee}{\end{eqnarray}}
\newcommand{\ba}{\begin{array}}
\newcommand{\ea}{\end{array}}
\newcommand{\bal}{\begin{align*}}
\newcommand{\eal}{\end{align*}}

\newcommand{\nn}{\nonumber}

\renewcommand{\(}{\Big(}
\renewcommand{\)}{\Big)}
\renewcommand{\[}{\Big[}
\renewcommand{\]}{\Big]}
\def \<{\langle}
\def \>{\rangle}
\definecolor{ggg}{rgb}{0,.6,0}
\begin{document}
~
\vspace{0.5cm}
\begin{center} {\Large \bf  Perturbative  OTOC  and Quantum Chaos in   Harmonic  Oscillators :   Second Quantization Method}
\\
(Revised)
                                                  
\vspace{1cm}

Wung-Hong Huang*\\
\vspace{0.5cm}
Department of Physics, National Cheng Kung University,\\
No.1, University Road, Tainan 701, Taiwan\\

\end{center}
\vspace{1cm}
\begin{center} {\large \bf  Abstract} \end{center}
 In this paper, the out-of-time-order correlators (OTOC) in  quantum harmonic oscillators are calculated analytically by second quantization method in  perturbative approximation.   We consider the coupled harmonic oscillators and anharmonic (quartic) oscillators.   The analytic formulas of microcanonical OTOC in the leading order  of interaction are obtained.  From these results we clearly see   that the interactions can enhance the correlation to very large values over time and is proportional to the energy level. The growth of OTOC signatures the  quantum chaos therein. The microcanonical OTOC which is an  increasing oscillation  becomes a simple increasing function in thermal OTOC after sum over all energy level, as  different level  oscillates with different frequency.  The property that, at late time stage OTOC  saturates to a constant value,  however, does not show in the first-order perturbation studied in this paper.
\\
\\
\\
\\
\scalebox{0.6}{\hspace{2.1cm}\includegraphics{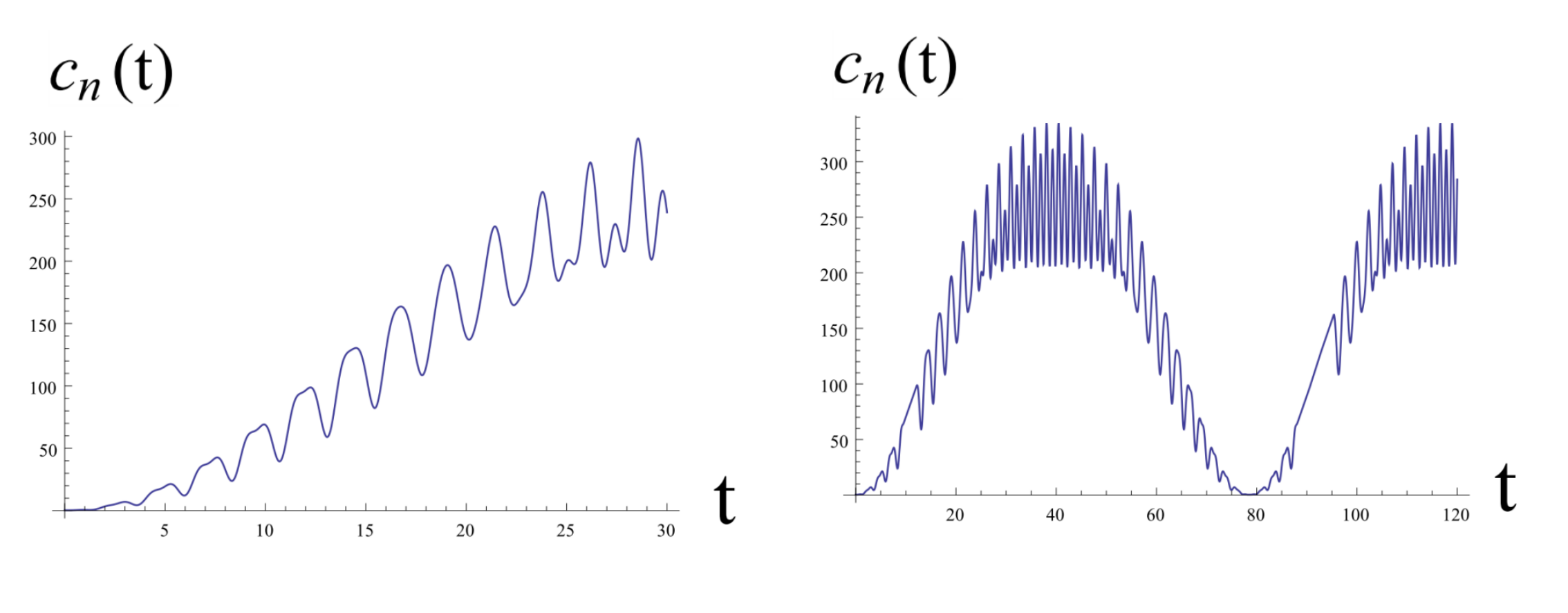}}
\\
\\
\begin{flushleft}
* Retired Professor of NCKU, Taiwan. \\
* E-mail: whhwung@mail.ncku.edu.tw
\end{flushleft}
%%%%%%%%%%%%%%%%%%%%%%%
%%%%%%%%%%%%%%%%%%%%%%%%%%%%%%%
\newpage
%%%%%%%%%%%%%%%%%%%%%%%
\tableofcontents
%%%%%%%%%%%%%%%%%%%%%%%
 %%%%%%%%%%%%%%%%%%%%%%%
\vspace{1cm}
%%%%%%%%%%%%%%%%%%%%%%%
 %%%%%%%%%%%%%%%%%%%%
%%%%%%%%%%%%%%%%%%%%%%%
%%%%%%%%%%%%%%%%%%%%
\section{Introduction}

The out-of-time-order correlator (OTOC) is defined by 
\be
C_T(t)=-\<[W(t),V(0)]^2\>_T
\ee
where the notation  $\< \cdot \>_T=Z^{-1}\text{tr}e^{-H/T}$ denotes the thermal expectation value at temperature T.  The quantity was first introduced to calculate a vertex correction of a current for a superconductor  \cite{Larkin}.  This quantum quantity  has attracted  a lot of attention in the physics community across many different fields, including quantum information, conformal field theory, AdS/CFT duality, and condensed matter physics  \cite{Kitaev15a,Kitaev15b,Shenker13a,Shenker14,Roberts14}.   

 Recently OTOC  is considered as a measure of the magnitude of quantum chaos.   The relation between the OTOC and chaos can be easily seen in the semiclassical limit if  W(t) = x(t)  and V = p.  In this case, we can  substitute  $f= x(t), g=p,   q_i= x(0)$, and $p_i= p$ into Poisson bracket to get $  \{f,g\}\equiv{\partial f\over \partial q_i }{\partial g\over\partial p_i }-{\partial f\over\partial p_i }{\partial g\over q_i }= {\partial x(t)\over \partial x(0)}$.   As the commutator $[x(t), p]$ becomes the Poisson bracket in the semiclassical limit we thus find that OTOC  $C(t)  =\hbar^2({\partial x(t)\over \partial x(0)})^2 $.  This gives the dependence of the final position on small changes in the initial position, the classical diagnostic of the butterfly effect \cite{Roberts14, Maldacena16}. 

The OTOC between a pair of local operators $W(t,x)$, $V(0 )$ can be expansed as   
\be
 C_T(t)
&=&  \<W(t)V(0)W(t)V(0)\>_T+\<V(0)W(t)V(0)W(t)\>_T    \nn\\
&& -\<W(t)V(0)V(0)W(t)\>_T-\<V(0)W(t)W(t)V(0) \>_T 
\ee
where the  last 2 terms are not directly sensitive to chaos  \cite{Shenker14}. For sufficiently late times, they are simply the time-independent disconnected diagram $\(\<W(t) W(t)\>_T\)\cdot\(\<V(0)V(0)\>_T\)$ while  last two terms are genuine out-of-time-order correlators and has an univeral behavior  
%%%%%%%%%%%%%%%%%%%%%%%%
\be
 & &{\<V(0)W(t)V(0)W(t)\>_T\over \<V(0)V(0))\>_T\<W(t)W(t)\>_T}=1-e^{\lambda_L(t-t_*-{|x|\over v_B})} 
\ee
in which
\\
$\bullet$ $t_*$ is a time scale called the scrambling time at which the commutator grows to be ${\cal O}(1)$. 
\\
$\bullet$ $v_B$ is the  buttery velocity   which characterizes speed at which the perturbation grows.  
\\
$\bullet$  $\lambda_L$  is the  Lyapunov exponent   which measures the rate of growth of chaos .
\\

The three parameters have been studied and calculated from gravity side \cite{Shenker13a, Shenker14, Roberts14, Maldacena16,  Shenker13b,Susskind,Liam,Verlinde} in which some interesting properties, such as  the Lyapunov exponent is bounded by temperature T : $\lambda \le  {2\pi  T}$, was found \cite{Maldacena16}. Along this way several literature  had studied the problem under external field or higher gravity theory, for example in \cite{Andrade,Sircar, Kundu,Ross,Huang16,Huang17,Huang18}.

The method of quantum theory to calculate OTOC  with general Hamiltonians was set up by Hashimoto recently in  \cite{Hashimoto17,Hashimoto20a,Hashimoto20b}.  For the simplest mode of the simple harmonic oscillator (SHO) the OTOC can be calculated exactly and  is purely oscillatory.  Some  complicated examples, such as the two-dimensional stadium billiard, could  exhibit classical chaos in which, after  numerical calculation shows that OTOCs   are growing non-exponentially at early times followed by a saturation at late times \cite{Hashimoto17}.  In an interesting paper \cite{Hashimoto20a}, the early-time exponential growth of OTOCs, which is expected to exhibit quantum chaos, was found in a system of non-linearly coupled oscillators (CHO) \cite{Hashimoto20b}.  The method has also been applied to study several systems  including many-body physics, for example in \cite{Das, Romatschke,Shen,Swingle-a, Cotler, Rozenbaum,Dymarsky,Bhattacharyya, Morita, Lin,Sundar,Swingle}.

In this framework  the properties of  OTOC  are mostly found by using numerical calculations.  The present paper is to study OTOC by analytic method under perturbation.  We try to see how the analytic  results from perturbative calculation could reveal the properties found in numerical calculation, at least qualitatively.  

In section II, we first review  Hashimoto's method  of  calculating quantum mechanic  OTOC  in SHO and then, use the second quantization method to obtain the same result quickly.

 In section III, we   use the   second quantization method to calculate OTOC in the systems  of  quantum coupled harmonic oscillator. We obtain the analytic result under  the perturbation.  Then,   we compare the results with those in \cite{Hashimoto20b}.  

 In section IV, we   use the   second quantization method to calculate OTOC in the systems  of  quantum anharmonic oscillator with  quartic interaction. We obtain the analytic result under  the perturbation.  Then we compare the results with those in \cite{Romatschke}.   Last section  is devoted to a short summary. 

%%%%%%%%%%%%%%%%%%%%%%%%
%%%%%%%%%%%%%%%%%%%%
%%%%%%%%%%%%%%%%%%%%%%%

%%%%%%%%%%%%%%%%%%%%
%%%%%%%%%%%%%%%%%%%%%%%%%%%%%%%%%%%%%%%%
\section{OTOC in Quantum Theory }
%%%%%%%%%%%%%%%%%%%%%%%
\subsection{Quantum Mechanic Approach to OTOC}
We first review the  Hashimoto's method of calculating  OTOC in quantum mechanic  model \cite{
 Hashimoto17}{\footnote{Sections 2.1 and 3.1 are used to review and simple extending Hashimoto method, so we use his notation $\hbar =1$. In other sections we restore $\hbar$ explicitly.}}.   For a general time-independent Hamiltonian: $H = H(x_1,....x_n,p_1,....p_n)$ the function of  OTOC is defined by
\be
C_T(t)=-\<[x(t),p(0)]^2 \>  
\ee
where $\<{\cal O)}\>\ \equiv tr(e^{-\beta H}\,{\cal O)})/tr e^{-\beta H}$.  Using energy eigenstates $|n\>$,  defined by
\be
H|n\>=E_n|n\>~~~~~~~~~\label{E}
\ee
 as the basis of the Hilbert space, then
\be
  C_T(t)={1\over Z}\sum_ne^{-\beta E_n}\,c_n(t) ,~~~c_n(t)\equiv -\<n|  [x(t),p(0)]^2         |n\>    \label{TC}
\ee
To calculate $c_n(t)$ we can inserte the  complete set $\sum_m|m\>\<m|=1$ to find an useful formula
\be
 c_n(t)  & =& -\sum_m\<n|  [x(t),p(0)]|m\>\<m|  [x(t),p(0)] |n\>  =  \sum_m(ib_{nm})(ib_{nm})^* ~\label{cn}\nn\\
b_{nm}&=& -i\<n|  [x(t),p(0)]|m\>,~~b_{nm}^*=b_{mn} 
\ee
To proceed, we substituting a relation $ x(t) = e^{ iHt }\,x\,e^{- iHt} $ and inserting the completeness relation again, and finally  obtain
\be
 b_{nm}&\equiv& -i\<n|  x(t), p(0)|m\> +i\<n| p(0) x(t),|m\>=-i\<n|  e^{ iHt }\,x\,e^{- iHt} p(0)|m\> +i\<n| p(0) e^{ iHt }\,x\,e^{- iHt}|m\>\nn\\
&=&-i\sum_k\<n|  e^{ iHt }\,x\,e^{- iHt}|k\>\<k| p(0)|m\>+i\sum_k<n| p(0) |k\>\<k|e^{ iHt }\,x\,e^{- iHt}|m\>  \nn\\
&=& -i\sum_k\(e^{iE_{nk}t}x_{nk}p_{km}-e^{iE_{km}t}p_{nk}x_{km}\)
\ee
where
\be
E_{nm}= E_n-E_m,~~~ x_{nm}=\<n| x|m\> ,~~p_{nm}=\<n|p |m\>~~~~\label{Enm}
\ee
\\
%%%%%%%%%%%%%%%%%%%%
For a natural Hamiltonian
\be
&&H=\sum_i{p_i^2\over 2m}+U(x_1,....x_N) 
\ee
by using the relations
\be
 [H,x_i]&=&-2ip_i\\
p_{km}&=&\<k|p |m\>={i\over 2}\<k|[H,x] |m\>={i\over 2}\<k|(H\,x)-(x\,H) |m\>\nn\\
&=&{i\over 2}\<k|(E_k\,x)-(x\,E_m) |m\>={i\over 2}(E_{km})x_{km} 
\ee
we have a simpe formula
\be
 b_{nm} & {=}& {1\over 2}\sum_k\,x_{nk}x_{km}\(e^{iE_{nk}t} E_{km} -e^{iE_{km}t}E_{nk}\) ~~~~~~~\label{bnm}
\ee
and  we can compute OTOCs through (\ref{bnm}) once we know $x_{nm}$ and $E_{nm}$ defined in (\ref{Enm}).

In the below, we first summarize the Hashimoto wavefunction method of  calculating  OTOC for SHO and then, present our method of second quantization to obtain the same result. 
%%%%%%%%%%%%%%%%%%%%%%%
%%%%%%%%%%%%%%%%%%%%%%%
\subsection{Simple Harmonic oscillator :   Wavefunction Method}
%%%%%%%%%%%%%%%%%%%%%%%
 For SHO model, which is an integrable example and no chaos, the Hamiltonian $H$, spectrum $E_n$ and state wavefunction $\Psi_n(x)$ have been shown in any textbook of quantum mechanics. 
%%%  https://en.wikipedia.org/wiki/Quantum_harmonic_oscillator          p=-i hbar \partial_x
\be
&& H={p^2\over 2M}+{M \omega^2\over2}x^2,~~~E_n=\hbar\omega\left(n+\frac{1}{2}\right),~~~E_{nm}=\hbar\omega(n-m)  \\
&&\Psi_n(x) = {1\over \sqrt{2^n\,n!}}\({M\omega\over \pi\hbar}   \)^{1/4}\,e^{-{M\omega x^2\over 2\hbar}}\,H_n\(\sqrt{M\omega\over \hbar}\,x\)~~~n =0,1,2,\cdots   
\ee
The functions $H_n$ are the physicists' Hermite polynomials.  Using the orthogonal and  recurrence relations of Hermite polynomials 
\be
\int dx H_m(x)H_n(x)e^{-x^2}&=&\sqrt \pi \ 2^n\ n! \ \delta_{nm}\\
H_{n+1}(x)&=&2x H_{n}(x)-2nH_{n-1}(x) 
% x H_{m}(x)&=&{1\over2}H_{m+1}(x)+mH_{m-1}(x)
\ee
we can   easily calculate $x_{nm}$.  The result  is  
\be
  x_{nm}&=&\int_{-\infty}^{\infty} dx \,\Psi^*_n(x)\,x\,\Psi_m(x)\\
&=& {1\over \sqrt{2^n\,n!}}{1\over \sqrt{2^m\,m!}}\({1\over \pi}   \)^{1/2}\,\int_{-\infty}^{\infty} dx e^{-{M\omega x^2\over  \hbar}}  H_n\(\sqrt{M\omega\over \hbar}\,x\)\ \sqrt{M\omega\over \hbar}\,x \     H_m\(\sqrt{M\omega\over \hbar}\,x\) \nn \\
 &=&{1\over \sqrt{2^n\,n!}}{1\over \sqrt{2^m\,m!}}\({1\over \pi}   \)^{1/2}\,\int_{-\infty}^{\infty} dx e^{-{M\omega x^2\over  \hbar}}  H_n\(\sqrt{M\omega\over \hbar}\,x\)\  \\
 &&~~~~~~~~~~~~~~~~~~~~~~~~~~~ ~~~~~\times   \[ {1\over2} H_{m+1}\(\sqrt{M\omega\over \hbar}\,x\) +mH_{m-1}\(\sqrt{M\omega\over \hbar}\,x\)\]    \\      
%&=&{1\over \sqrt{2^n\,n!}}{1\over \sqrt{2^m\,m!}}\({1\over \pi}   \)^{1/2}\,\sqrt{\hbar\over M\omega}\,%\int_{-\infty}^{\infty} dy e^{-y^2 }  H_n(y) \[ {1\over2} H_{m+1}(y)+mH_{m-1}(y)\]    \\     
%&=&{1\over \sqrt{2^n\,n!}}{1\over \sqrt{2^m\,m!}}\({1\over \pi}   \)^{1/2}\,\sqrt{\hbar\over M\omega}\  \  \sqrt \pi \ 2^n\ n! \   \[ {1\over2} \delta_{n(m+1)}+m\delta_{n (m-1)}\]    \\                      
%\\
%&=& \sqrt{\hbar\over M\omega}\[{1\over \sqrt{2^n\,n!}}{1\over \sqrt{2^m\,m!}}   \ 2^n\ n! \    {1\over2} \delta_{n,m+1}+{1\over \sqrt{2^n\,n!}}{1\over \sqrt{2^m\,m!}}   \ 2^n\ n! \ \ m\delta_{n,m-1}\]    \\                      
%\\
%&=& \sqrt{\hbar\over  M\omega}\, \(\sqrt{m+1\over 2}  \ \delta_{n,m+1}+\sqrt{m\over 2}  \ \delta_{n,m-1}\), ~~~~~~n,m=0,1,2,\cdots\\
&=& \sqrt{\hbar\over  2M\omega}\, \(\sqrt{n }  \ \delta_{n,m+1}+\sqrt{n+1 }  \ \delta_{n,m-1}\), ~~~~~~n,m=0,1,2,\cdots~~~~~~\label{xnm}
  \ee
Substituting above expressions into  (\ref{Enm}) and  (\ref{bnm}) we obtain 
\be 
 b_{nm}(t)&=&{1\over 2}\sum_k\,x_{nk}x_{km}\(e^{iE_{nk}t} E_{km} -e^{iE_{km}t}E_{nk}\) \\
& {=}&  \( -\frac{\hbar^2 n \cos (\hbar \omega t)}{2 M}+  \frac{\hbar^2 \sqrt{n+1} \sqrt{n+1} \cos (\hbar \omega t )}{2 M}  \)\ \delta_{m,n}         \label{first}         \\
&=&\frac{\hbar^2   }{2 M}  \cos(\hbar \omega t)\ \delta_{nm}
\ee
in which the first term in (\ref{first}) is coming from k=n-1 while second term is from k=n+1. Then
\be
c_n(t)&=&\(\frac{\hbar^2   }{2 M} \)^2\cos^2 (\hbar \omega t),~~~C_T(t)=\(\frac{\hbar^2   }{2 M} \)^2\cos^2(\hbar \omega t) ~~~~~~~~~~~\label{HR}
\ee
It is seen that both of $c_T(t)$ and $C_T(t)$ are periodic functions.  They do not depend on energy level $n$ or temperature $T$. This is a special property only for the harmonic oscillator among  several examples.
%%%%%%%%%%%%%%%%%%%%
\subsection{Simple Harmonic oscillator :  Second Quantization Method}
In the second quantization the states denoted as $|n\>$ are   created and destroyed by the operatrors $a^\dag$ and $a$ respectively. There are following basic properties
\be
&&[a,a^\dag]=1,~~~[a,a]= [a^\dag,a^\dag]=0,~~~~\<n|m\>=\delta_{m,n}\\
&&a^\dag|n\>= \sqrt{n+1} |n+1\>,~~~  a|n\>= \sqrt{n} |n-1\>,~~~a^\dag a |n\>=n |n \>
\ee
%%%  https://en.wikipedia.org/wiki/Quantum_harmonic_oscillator          p=-i hbar \partial_x
Applying above  relations and following definitions
\be
&&x=\sqrt{\hbar\over2M\omega}(a^\dag+a ),~~~p= i\sqrt{M\omega\hbar\over2}(a^\dag-a) 
\ee
to SHO, then, the Hamiltonian $H$, spectrum $E_n$ and $E_{nm}$ become 
\be
  H&=&{p^2\over 2M}+{M\omega^2\over2}x^2=-{\omega\hbar\over4 }(a- a^\dag)^2+{\omega\hbar\over4 }(a+ a^\dag)^2\nn\\
 &=&{\hbar\omega\over2 } \(   a a^\dag +   a^\dag a   \)   = \hbar\omega \( a^\dag a+ \frac{1}{2}\) \\
H|n\>&=&E_n|n\>,~~E_n=\hbar\omega\left(n+\frac{1}{2}\right), ~~~E_{nm}=\hbar\omega(n-m) 
\ee
Basic  relations
\be
x|n\>=\sqrt{\hbar\over2M\omega}(a^\dag+a ) |n\>=\sqrt{\hbar\over2M\omega}\ \sqrt{n}|n-1\>+  \sqrt{\hbar\over2M\omega}\ \sqrt{n+1} |n+1\>
\ee
quickly leads to
\be
x_{nm}&\equiv&\<n|x|m\>=\sqrt{\hbar\over2M\omega}\( \sqrt{m}\ \delta_{n,m-1}+   \sqrt{m+1}\ \delta_{n, m+1}\)   \label{SQxnm}
\ee
which exactly reproduces (\ref {xnm}) and, therefore the values of  $b_{nm}$, $c_n(t)$ and $C_T(t)$ in (\ref{HR}).   Note that the extending to the  supersymmetric quantum  harmonic oscillator was studied in \cite{Das}.
%
%%%%%%%%%%%%%%%%%%%%
%%%%%%%%%%%%%%%%%%%%%%%%% 
 \section{Perturbative OTOC of Coupled Harmonic Oscillators}
%%%%%%%%%%%%%%%%%
\subsection{Formulas of OTOC  in Two Dimension  }
For theory on two dimensions with coordinate (x,y) we can generalize  the formula of Hashimoto \cite{Hashimoto17} to calculate the associated OTOCs. Denote $x_1=x, x_2=y, p_1=p_x, p_2=p_y, \vec n=(n_1,n_2) =(n_x,n_y)$ then
\be
  C_T^{ij }(t)={1\over Z}\sum_{{\vec n}} e^{-\beta E_{\vec n}}\,c^{ij }_{{\vec n}}(t) ,~~~c^{i j}_{{\vec n}}(t)\equiv -\<{\vec n}|  [x_i(t),p_j(0)]^2         |{\vec n}\> 
\ee
To calculate $c^{ij }_{{\vec n}}(t)$ we can inserte the  complete set $\sum_{\vec m}|{\vec m}\>\<{\vec m}|=1$ to find an useful formula
\be
c^{i j}_{{\vec n}}(t) & =& -\sum_{\vec m}\<{\vec n}|  [x_i(t),p_j(0)]|{\vec m}\>\<m|  [x_i(t),p_j(0)] |n\>  =  \sum_m(\text{i}b^{i j}_{\vec n\vec m})(\text{i}b^{i j}_{\vec n\vec m})^* ~\label{cij}\nn\\
b^{i j}_{\vec n\vec m}&=& -\text{i}\<\vec n|  [x_i(t),p_j(0)]|\vec m\>,~~(b^{i j}_{\vec n\vec m})^*=b^{i j}_{\vec m\vec n} 
\ee
To proceed, we substituting a relation $ x_i(t) = e^{ iHt }\,x_i\,e^{- iHt} $ and inserting the completeness relation again, and finally  obtain
\be
 b^{i j}_{\vec n\vec m}&\equiv& \text{-i}\<\vec n|  x_i(t), p_j(0)|\vec m\> +\text{i}\<\vec n| p_j(0) x_i(t),|\vec m\>=-\text{i}\<\vec n|  e^{ iHt }\,x_i\,e^{- iHt} p_j(0)|\vec m\> +\text{i}\<\vec n| p_j(0) e^{ iHt }\,x_i\,e^{- iHt}|\vec m\>\nn\\
&=&\text{-i}\sum_{\vec k}\<\vec n|  e^{ iHt }\,x_i\,e^{- iHt}|\vec k\>\<\vec k| p_j(0)|\vec m\>+\text{i}\sum_{\vec k}<\vec n| p_j(0) |\vec k\>\<\vec k|e^{ iHt }\,x_i\,e^{- iHt}|\vec m\>  \nn\\
&=& \text{-i}\sum_{\vec k}\(e^{iE_{\vec n\vec k}t}x^i_{\vec n\vec k}p^j_{\vec k\vec m}-e^{iE_{\vec k\vec m}t}p^j_{\vec n\vec k}x^i_{\vec k\vec m}\)                
\ee
where
\be
E_{\vec n\vec m}= E_{\vec n}-E_{\vec m},~~~ x^i_{\vec n\vec m}=\<\vec n| x^i|\vec m\> ,~~p^i_{\vec n\vec m}=\<\vec n|p^i |\vec m\> 
\ee
\\
%%%%%%%%%%%%%%%%%%%%
For a natural Hamiltonian
\be
&&H=\sum_i{p_i^2\over 2m}+U(x_1,....x_N) 
\ee
by using the relations
\be
 [H,x_i]&=&-2\text{i}p_i\\
p^i_{\vec k\vec m}&=&\<\vec k|p_i |\vec m\>={\text{i}\over 2}\<\vec k|[H,x_i] |\vec m\>={\text{i}\over 2}\<\vec k|(H\,x_i)-(x_i\,H) |\vec m\>\nn\\
&=&{\text{i}\over 2}\<\vec k|(E_{\vec k}\,x_i)-(x_i\,E_{\vec m}) |\vec m\>={\text{i}\over 2}(E_{\vec  k\vec m})x^i_{\vec  k\vec  m} 
\ee
we have a simple formula
\be
 b^{i j}_{\vec n\vec m} & {=}& {1\over 2}\sum_{\vec k}\,x^i_{\vec n\vec k}x^j_{\vec k\vec m}\(e^{iE_{\vec n\vec k}t} E_{\vec k\vec m} -e^{iE_{\vec k\vec m}t}E_{\vec n\vec k}\) ~~~~~~~\label{cbij}
\ee
which is the two-dimension extension of  (\ref{bnm}) and  we can use it to compute OTOCs, $c^{i j}_{{\vec n}}(t)$,  once we know $x^i_{\vec n  \vec k  }$ and $E_{\vec n  \vec k  }$.
%%%%%%%%%%%%%%%%%%%%%%%%%
\subsection{Coupled Simple Harmonic Oscillators}
Use basic relations
\be
&&x=\sqrt{\hbar\over2M\omega}(a_x^\dag+a_x ),~~~p_x= i\sqrt{M\omega\hbar\over2}(a_x^\dag-a_x) \\
&&y=\sqrt{\hbar\over2M\omega}(a_y^\dag+a_y ),~~~p_y= i\sqrt{M\omega\hbar\over2}(a_y^\dag-a_y) 
\ee
 the Hamiltonian $H$ for a system of two harmonic oscillators coupled nonlinearly with each other descrided in  \cite{Hashimoto20b} becomes
\be
  \tilde H&=&{p_x^2\over 2M}+{M\omega^2\over2}x^2+{p_y^2\over 2M}+{M\omega^2\over2}y^2+gx^2y^2\\
 &=& \hbar\omega \( a_x^\dag a_x+a_y^\dag a_y+ 1\) +g\({\hbar\over2M\omega}\)^2(a_x^\dag+a_x )^2(a_y^\dag+a_y )^2 \\
&=&H +V
\ee
% %%%%%%%%%% %%%%%%%%%%
which has  a well-known unperturbed solution
\be
H |\vec n\>&=&E_{\vec n}|n_x,n_y\>,~~E_{\vec n}=\hbar\omega (n_x+n_y+1 )\\
E_{\vec n \vec m}&=&\hbar\omega(n_x+n_y-m_x-m_y ) \\
x_{\vec n\vec m}=x(\text{n}_x,\text{n}_y,\text{m}_x,\text{m}_y)&=& \frac{\sqrt{\frac{\hbar }{M \omega }} \left(\sqrt{\text{m}_x} \delta _{\text{m}_x-1,\text{n}_x}+\sqrt{\text{m}_x+1} \delta _{\text{m}_x+1,\text{n}_x}\right)}{\sqrt{2}}\  \delta _{\text{m}_y,\text{n}_y}           \label{CSQxija}\\
y_{\vec n\vec m}=y(\text{n}_x,\text{n}_y,\text{m}_x,\text{m}_y)&=&  \frac{\sqrt{\frac{\hbar }{M \omega }} \left(\sqrt{\text{m}_y} \delta _{\text{m}_y-1,\text{n}_y}+\sqrt{\text{m}_y+1} \delta _{\text{m}_y+1,\text{n}_y}\right)}{\sqrt{2}}\  \delta _{\text{m}_x,\text{n}_x}   \label{CSQxijb}
\ee
The first-order perturbation formulas in quantum mechanics textbook are
\be
\widetilde {E_{\vec n}}&\approx& E_{\vec n} +\<\vec n |\, V  \,   |\vec n  \>   \\
\widetilde {|\vec n\>}&\approx& |\vec n \>+\sum_{\{\vec  n\ne \vec m\}'}{|\vec m \>\<\vec m |  V    |\vec n \> \over E_{\vec n} -E_{\vec m} } \>        \label{sumE}
\ee
where $\{\vec  n\ne \vec m\}'$ denotes   all the states with  $ E_{\vec n} \ne E_{\vec m} $, not just $\vec  n\ne \vec m$.  In our study there are states of $\vec  n\ne \vec m$ while $ E_{\vec n} = E_{\vec m} $ and they are excluded in the above summation.
%%%%%%%%%%%%%%%%
 
A basic quantity we need, after calculation, is
\be
V|n_x,n_y\>&=&\frac{g \hbar ^4}{16 M^4 \omega ^4}\(\sqrt{n_x -1 } \sqrt{n_x} \sqrt{n_y-1} \sqrt{n_y}~|n_x-2,n_y-2\> +\sqrt{n_x-1} \sqrt{n_x} \left(2 n_y+1\right)~|n_x-2,n_y\>  \nn\\
&&+\sqrt{n_x-1} \sqrt{n_x} \sqrt{n_y+1} \sqrt{n_y+2}~|n_x-2,n_y+2\>   \nn\\
&&+\left(2 n_x+1\right) \sqrt{n_y-1} \sqrt{n_y}~|n_x,n_y-2\>        +\left(2 n_x+1\right) \left(2 n_y+1\right)~|n_x,n_y\>\nn\\
&&+\left(2 n_x+1\right) \sqrt{n_y+1} \sqrt{n_y+2}~|n_x,n_y+2\>              \nn    \\
&&   +\sqrt{n_x+1} \sqrt{n_x+2} \sqrt{n_y-1} \sqrt{n_y}~|n_x+2,n_y-2\>+\sqrt{n_x+1} \sqrt{n_x+2} \left(2 n_y+1\right)~|n_x+2,n_y\>       \nn  \\
&&    +\sqrt{n_x+1} \sqrt{n_x+2} \sqrt{n_y+1} \sqrt{n_y+2}~|n_x+2,n_y+2\>\)
\ee
% %%%%%%%%%%%%%%%%%% 
The first-order energy $ \widetilde {E_{\vec n}}$ and Fock state $ \widetilde { |\vec n  \>} $ become 
\be
 \widetilde {E_{n_x,n_y}} &\approx&  E_{n_x,n_y} +\<n_x,n_y |V   |n_x,n_y  \>=\hbar\omega (n_x+n_y+1 )+\frac{g \hbar ^4}{16 M^4 \omega ^4} \left(2 n_x+1\right) \left(2 n_y+1\right)+ {\cal O}(g^2) ~~\label{cE1}  \\
\nn\\
\widetilde { |n_x,n_y  \>}&\approx&   |n_x,n_y  \> + \frac{g \hbar^3  \sqrt{n_x-1} \sqrt{n_x} \sqrt{n_y-1} \sqrt{n_y} }{64 M^4 \omega ^5}~|n_x-2,n_y-2 \>    \nn\\
\nn\\
&&  +\frac{g \hbar^3  \sqrt{n_x-1} \sqrt{n_x} \left(2 n_y+1\right) }{32 M^4 \omega ^5}~|n_x-2,n_y \>+\frac{g \hbar^3  \left(2 n_x+1\right) \sqrt{n_y-1} \sqrt{n_y} }{32 M^4 \omega ^5}                  ~|n_x,n_y-2 \> \nn\\
\nn\\
&& -\frac{g \hbar^3  \left(2 n_x+1\right) \sqrt{n_y+1} \sqrt{n_y+2} }{32 M^4 \omega ^5}~|n_x,n_y+2 \> -\frac{g \hbar^3  \sqrt{n_x+1} \sqrt{n_x+2} \left(2 n_y+1\right) }{32 M^4 \omega ^5}    ~|n_x+2,n_y \> \nn\\
\nn\\
&&  -\frac{g \hbar^3  \sqrt{n_x+1} \sqrt{n_x+2} \sqrt{n_y+1} \sqrt{n_y+2}}{64 M^4 \omega ^5}  ~|n_x+2,n_y+2 \>+ {\cal O}(g^2) 
\ee
in which the states $|n_x-2,n_y+2 \>$ and $ |n_x+2,n_y-2 \>$  doe not appear on right-hand side because $E_{n_x-2,n_y+2}=E_{n_x-2,n_y+2}=E_{n_x,n_y}$, as mentioned in (\ref{sumE}).
%%%%%%%%%%%%%%%%%%%%%%%%%
\subsection{First-order Microcanonical OTOC of CHO}
%%%%%%%%%%%%%%%%%% 
To proceed, we have  an interesting and simple relation
\be
&& x  \widetilde{ |n_x,n_y  \>}\nn\\
&=& \xi _{-1,-2}(n_x,n_y)\  \widetilde{ |n_x-1,n_y-2  \>}+\xi _{-1,0}(n_x,n_y)\  \widetilde{ |n_x-1,n_y  \>}\nn\\
&&+\xi _{-1,2}(n_x,n_y)\  \widetilde{ |n_x-1,n_y+2  \>}\nn\\
&&+\xi _{1,-2}(n_x,n_y)\  \widetilde{ |n_x+1,n_y-2  \>}+\xi _{1,0}(n_x,n_y)\  \widetilde{ |n_x+1,n_y  \>} \nn\\
&&  +\xi _{1,2}(n_x,n_y) \  \widetilde{ |n_x+1,n_y+2  \>}  + {\cal O}(g^2) 
\ee
in which  the six coefficients $\xi _{i_x,i_y}(n_x,n_y)$ are
\be
\xi _{-1,-2}(n_x,n_y)&=& \frac{g \sqrt{\text{n}_x} \sqrt{\text{n}_y-1} \sqrt{\text{n}_y} \ \hbar ^3 \sqrt{\frac{\hbar }{M \omega }}}{32 \sqrt{2} M^4 \omega ^5}                  \\
\xi _{-1,0}(n_x,n_y)&=&  \frac{\sqrt{\text{n}_x} \sqrt{\frac{\hbar }{M \omega }} \left(16 M^4 \omega ^5-g (2 \text{n}_y+1) \ \hbar ^3\right)}{16 \sqrt{2} M^4 \omega ^5}                   \\
\xi _{-1,2}(n_x,n_y)&=&   -\frac{g \sqrt{\text{n}_x} \sqrt{\text{n}_y+1} \sqrt{\text{n}_y+2} \ \hbar ^3 \sqrt{\frac{\hbar }{M \omega }}}{16 \sqrt{2} M^4 \omega ^5}        \\
\nn\\
\xi _{1,-2}(n_x,n_y)&=&   -\frac{g \sqrt{\text{n}_x+1} \sqrt{\text{n}_y-1} \sqrt{\text{n}_y} \ \hbar ^3 \sqrt{\frac{\hbar }{M \omega }}}{16 \sqrt{2} M^4 \omega ^5}         \\
\xi _{1,0}(n_x,n_y)&=&  \frac{\sqrt{\text{n}_x+1} \sqrt{\frac{\hbar }{M \omega }} \left(16 M^4 \omega ^5-g (2 \text{n}_y+1) \ \hbar ^3\right)}{16 \sqrt{2} M^4 \omega ^5}                \\
\xi _{1,2}(n_x,n_y)&=&  \frac{g \sqrt{\text{n}_x+1} \sqrt{\text{n}_y+1} \sqrt{\text{n}_y+2} \ \hbar ^3 \sqrt{\frac{\hbar }{M \omega }}}{32 \sqrt{2} M^4 \omega ^5}                
  \ee
For $g=0$ the coefficient $\xi _{i_x,i_y}$ remains only two non-zero components $\xi _{-1,0},~\xi _{1,0}$ which exactly reproduce the function in  SHO (\ref{SQxnm}). 

 Above result leads to 
\be
\widetilde{{x}_{\vec n\vec m}}&=&\widetilde{\< n_x,n_y |}   x  \widetilde{ |m_x,m_y  \>}=\sum_{i_x ,i_y}\ \xi _{ i_x,i_y}(m_x,m_y)\ \delta_{m_x+i_x,n_x,}\delta_{m_y+i_y,n_y,}     ~\label{cxnm}\\
&=&\ \xi _{-1,-2}(m_x,m_y)  \   \delta _{m_x-1,n_x}  \delta _{m_y-2,n_y} + \xi _{-1,0}(m_x,m_y)  \   \delta _{m_x-1,n_x}\  \delta _{m_y ,n_y}  \nn\\
&&\ + \xi _{-1,2}(m_x,m_y)  \   \delta _{m_x-1,n_x}  \delta _{m_y+2 ,n_y}              \nn \\
&& +\ \xi _{ 1,-2}(m_x,m_y)  \   \delta _{m_x+1,n_x}  \delta _{m_y-2,n_y} + \xi _{ 1,0}  \   \delta _{m_x+1,n_x}(m_x,m_y) \ \delta _{m_y ,n_y} \nn\\
&& \  + \xi _{ 1,2}(m_x,m_y)  \   \delta _{m_x+1,n_x}  \delta _{m_y+2 ,n_y}   
\ee
We   substitute spectrum   $\widetilde{{E}_{\vec n}}$ in (\ref{cE1}) and  matrix element  $\widetilde{{x}_{\vec n\vec m}}$ in (\ref{cxnm}) into formula   (\ref{cbij2}), which is the new notation of (\ref{cbij}),  
\be 
b^{xx}_{\vec n\vec m} & {=}& {1\over 2}\sum_{\vec k}\,\widetilde{x _{\vec n\vec k}}\widetilde{x _{\vec k\vec m}}\(e^{i\widetilde{E_{\vec n\vec k}}t} \widetilde{E_{\vec k\vec m}} -e^{i\widetilde{E_{\vec k\vec m}}t}\widetilde{E_{\vec n\vec k}}\) ~~~~~~~\label{cbij2}  
\ee
to find $ b^{xx}_{\vec n\vec m}$, which can be expressed as a summation of following   six terms : 
\be
&&{1\over 2} \widetilde{x _{\vec n\vec k}}\widetilde{x _{\vec k\vec m}}\(e^{i\widetilde{E_{\vec n\vec k}}t} \widetilde{E_{\vec k\vec m}} -e^{i\widetilde{E_{\vec k\vec m}}t}\widetilde{E_{\vec n\vec k}}\)   \nn\\
&=& \zeta_{-1,-2}(\vec n,\vec m)\ \delta_{k_x,n_x-1} \delta_{k_y,n_y-2}+  \zeta_{-1,0}(\vec n,\vec m)\ \delta_{k_x,n_x-1}  \delta_{k_y,n_y}+ \zeta_{-1, 2}(\vec n,\vec m)\ \delta_{k_x,n_x-1}  \delta_{k_y,n_y+2}                               \nn \\
&&+ \zeta_{ 1,-2}(\vec n,\vec m)\ \delta_{k_x,n_x+1}  \delta_{k_y,n_y-2}+  \zeta_{ 1,0}(\vec n,\vec m)\ \delta_{k_x,n_x+1}  \delta_{k_y,n_y}+ \zeta_{ 1, 2}(\vec n,\vec m)\ \delta_{k_x,n_x+1}  \delta_{k_y,n_y+2}\nn\\                               
\ee
which are used to study the OTOC $\<[x(t),p(0)]^2\>$. The quantity OTOC $\<[y(t),p(0)]^2>$ could be obtain by exchanging $x\leftrightarrow y$ due to the symmetry of nonlinear interaction $g x^2y^2$ in CHO.

Explicit forms of the six coefficients can be obtained which  are expressed as
%%%%%%%%%%%%%%%        %%%%%%%%%%%%%%%
\be
\zeta_{-1,-2}(\vec n,\vec m)&=&\eta_{-1,-2}^{0,2}\  
\delta_{\text{m}_x,\text{n}_x}\delta_{\text{m}_y,\text{n}_y+2}+\eta_{-1,-2}^{- 2, 2}\  
 \delta_{\text{m}_x+2,\text{n}_x}\delta_{\text{m}_y,\text{n}_y+2}  \\
\zeta_{-1,0}(\vec n,\vec m)&=& \eta_{-1,0}^{ 0,2}\    \delta_{\text{m}_x ,\text{n}_x}\delta_{\text{m}_y,\text{n}_y+2}+ \eta_{-1,0}^{-2  , 2  }\ \delta_{\text{m}_x+2,\text{n}_x}\delta_{\text{m}_y,\text{n}_y+2}+ \eta_{-1,0}^{0  , -2 }\ \delta_{\text{m}_x,\text{n}_x}\delta_{\text{m}_y+2,\text{n}_y}\nn\\
&&+ \eta_{-1,0}^{ -2 ,-2  }\delta_{\text{m}_x+2,\text{n}_x}\delta_{\text{m}_y+2,\text{n}_y}+ \eta_{-1,0}^{ 0 , 0 } \delta_{\text{m}_x,\text{n}_x}\delta_{\text{m}_y,\text{n}_y}   \\
\zeta_{-1, 2}(\vec n,\vec m)&=&\eta_{-1, -2}^{0,-2}\  
 \delta_{\text{m}_x,\text{n}_x}\delta_{\text{m}_y+2,\text{n}_y}+\eta_{-1, 2}^{- 2, -2}\ \delta_{\text{m}_x+2,\text{n}_x }\delta_{ \text{m}_y+2,\text{n}_y} 
\\
\nn\\
  \zeta_{ 1,-2}(\vec n,\vec m)&=&\eta_{ 1,-2}^{2,2}\  
\delta_{\text{m}_x,\text{n}_x+2}\delta_{\text{m}_y,\text{n}_y+2}+\eta_{ 1,-2}^{0, 2}\  
 \delta_{\text{m}_x,\text{n}_x}\delta_{ \text{m}_y,\text{n}_y+2}  \\
\zeta_{ 1,0}(\vec n,\vec m)&=& \eta_{ 1,0}^{ 2,-2    }\  \delta_{\text{m}_x,\text{n}_x+2}\delta_{  \text{m}_y+2,\text{n}_y}+ \eta_{ 1,0}^{0  , -2  }\delta_{\text{m}_x,\text{n}_x }\ \delta_{ \text{m}_y+2,\text{n}_y }+ \eta_{ 1,0}^{2  , 2 }\ \delta_{\text{m}_x,\text{n}_x+2}\delta_{\text{m}_y,\text{n}_y+2}\nn\\
&&+ \eta_{ 1,0}^{ 0 , 2  }\  \delta_{\text{m}_x,\text{n}_x}\delta_{\text{m}_y,\text{n}_y+2}+ \eta_{ 1,0}^{ 0 , 0 }\  \delta_{\text{m}_x,\text{n}_x}\delta_{\text{m}_y,\text{n}_y}   \\
\zeta_{ 1, 2}(\vec n,\vec m)&=&\eta_{ 1,  2}^{2,-2}\  
 \delta_{\text{m}_x,\text{n}_x+2}\delta_{\text{m}_y+2,\text{n}_y}+\eta_{ 1, 2}^{0,  2}\ \delta_{\text{m}_x,\text{n}_x}\delta_{\text{m}_y+2,\text{n}_y} 
\ee
where, for example 
\be
\eta_{-1,-2}^{0,2}&=&-\frac{g \sqrt{\text{m}_y} \text{n}_x \sqrt{\text{n}_y+1} \ \hbar ^5  \(1+3 e^{\frac{i g t \hbar ^4 (\text{n}_x+\text{n}_y+1)}{2 M^4 \omega ^4}+4 i t \omega  \hbar } \) \ e^{ -\frac{i g t \hbar ^4 (4 \text{n}_x+2 \text{n}_y+3)}{8 M^4 \omega ^4}-3 i t \omega  \hbar  }}{128 M^5 \omega ^5}    \\
\nn\\
\eta_{-1,-2}^{ -2, 2}&=& \frac{g \sqrt{\text{m}_x+1} \sqrt{\text{m}_y} \sqrt{\text{n}_x} \sqrt{\text{n}_y+1}\ \hbar ^5 \(1+e^{  \frac{i g (2 \text{n}_x-3) t \hbar ^4}{4 M^4 \omega ^4}+2 i t \omega  \hbar }\)  \ e^{ -\frac{i g t \hbar ^4 (4 \text{n}_x-2 \text{n}_y-7)}{8 M^4 \omega ^4}-  i t \omega  \hbar  }}{64 M^5 \omega ^5}           \nn \\
\\
\eta_{ 1,0}^{0,-2}&=&\frac{g \sqrt{\text{m}_y+1} (\text{n}_x+1) \sqrt{\text{n}_y}\ \hbar ^5 \left(-1+e^{\frac{i g t \hbar ^4 (\text{n}_x-\text{n}_y+2)}{2 M^4 \omega ^4}}\right)\ e^{\frac{i g (2 \text{ny}-3) t \hbar ^4}{8 M^4 \omega ^4}+i t \omega  \hbar }}{64 M^5 \omega ^5}
\ee
 After summing up  $\vec k$ the function $ b^{xx}_{\vec n\vec m}$ in (\ref{cbij2}) can be expressed as a summation of following   seven  terms 
\be
b^{xx}_{\vec n\vec m}&=& b_{-2,-2}\ \delta_{m_x,n_x-2}\ \delta_{m_y,n_y-2}+ b_{-2, 2}\ \delta_{m_x,n_x+2}\ \delta_{m_y,n_y-2}\nn\\
&&+ b_{0,-2}\ \delta_{m_x,n_x}\ \delta_{m_y,n_y-2}+ b_{0, 0}\ \delta_{m_x,n_x }\ \delta_{m_y,n_y }+ b_{0, 2}\ \delta_{m_x,n_x }\ \delta_{m_y,n_y +2}\nn\\
&&+ b_{ 2,-2}\ \delta_{m_x,n_x+2}\ \delta_{m_y,n_y-2}+ b_{ 2, 2}\ \delta_{m_x,n_x+2}\  \delta_{m_y,n_y+2}    \label{Cbxx}
\ee
where 
\be
 b_{-2,-2}&=& \frac{1}{128 M^5 \omega ^5}\ e^{\frac{i g (2 \text{n}_y-5) t \hbar ^4}{8 M^4 \omega ^4}+i t \omega  \hbar }\ \(-1+e^{\frac{i g t \hbar ^4}{2 M^4 \omega ^4}}\)                  \nn\\
&&\times \(e^{\frac{i g \text{n}_x t \hbar ^4}{2 M^4 \omega ^4}+2 i t \omega  \hbar }+3 e^{\frac{i g t \hbar ^4}{4 M^4 \omega ^4}}\)\ g \sqrt{\text{n}_x-1} \sqrt{\text{n}_x} \sqrt{\text{n}_y+1} \sqrt{\text{n}_y+2} \hbar ^5\\ 
 b_{-2, 2}&=& \frac{1}{64 M^5 \omega ^5}\ e^{-\frac{i g (4 \text{n}_x-2 \text{n}_y-1)) t \hbar ^4}{8 M^4 \omega ^4}-i t \omega  \hbar }\  \(-1+e^{\frac{i g t \hbar ^4}{2 M^4 \omega ^4}}\)                  \nn\\
&&\times \(e^{\frac{i g t \hbar ^4}{4 M^4 \omega ^4}}-e^{\frac{i g \text{n}_x t \hbar ^4}{2 M^4 \omega ^4}+2 i t \omega  \hbar }\)\ g \sqrt{\text{n}_x-1} \sqrt{\text{n}_x} \sqrt{\text{n}_y+1} \sqrt{\text{n}_y+2} \hbar ^5 \\
 b_{0, 0}&=&  \frac{\hbar ^2}{2 M}{ \cos \left(\frac{g (2 \text{n}_y+1) t \hbar ^4}{8 M^4 \omega ^4}+t \omega  \hbar \right)}\\
 b_{2, -2}&=&-\frac{1}{64 M^5 \omega ^5} {\left(-1+e^{\frac{i g t \hbar ^4}{2 M^4 \omega ^4}}\right) \exp \left(-   \left(\frac{i t  g (2 \text{n}_y +1) \hbar ^4}{8 \omega ^4M^4} +  i t \omega  \hbar\right) \right)} \nn\\
&&      g \sqrt{\text{n}_x +1} \sqrt{\text{n}_x +2} \sqrt{\text{n}_y -1} \sqrt{\text{n}_y } \hbar ^5 \left(-1+\exp \left(\frac{i g t \hbar ^4 (2 \text{n}_x +3) }{4 M^4 \omega ^4}+2 i t \omega  \hbar \right)\right)                      \\
b_{2,  2}&=&-\frac{1}{128 M^5 \omega ^5}{\left(-1+e^{\frac{i g t \hbar ^4}{2 M^4 \omega ^4}}\right) \exp \left(-\frac{i g t \hbar ^4 (4 \text{n}_x +2 \text{n}_y +11)}{8 M^4 \omega ^4}-3 i t \omega  \hbar \right)}  \nn\\
&& g \sqrt{\text{n}_x +1} \sqrt{\text{n}_x +2} \sqrt{\text{n}_y +1} \sqrt{\text{n}_y +2} \hbar ^5 \left(1+3 \exp \left(\frac{i g (2 \text{n}_x +3) t \hbar ^4}{4 M^4 \omega ^4}+2 i t \omega  \hbar \right)\right)
\ee
while $b_{0,-2}$ and $b_{0, 2} $  are  given in appendix   A. 

Collect all we finally obtain  the analytic formula of microcanonical OTOC of coupled harmonic oscillator
\be
c^{xx}_{\vec n}(t)&=&b_{-2,-2}b_{-2,-2}^*+b_{-2, 2}b_{-2, 2}^*+b_{0,-2}b_{0,-2}^*+b_{0, 0}b_{0, 0}^*+b_{0, 2}b_{0 , 2}^*+  b_{ 2,-2}b_{ 2,-2}^*+b_{ 2, 2}b_{-2, 2}^*      \label{Ccn}\nn\\
\ee
in which  
\be
&&b_{-2, -2}b_{-2,- 2}^*=\frac{g^2 (\text{n}_x-1) \text{n}_x \left(\text{n}_y^2+3 \text{n}_y+2\right) \hbar ^{10} \sin ^2\left(\frac{g t \hbar ^4}{4 M^4 \omega ^4}\right)\( 5+3 \cos \left(\frac{t \hbar g (2 \text{n}_x-1) \hbar ^3  }{4 M^4 \omega ^4}+2t \omega  \hbar\right)\)}{2048 M^{10} \omega ^{10}} \nn\\
 \nn\\
&&b_{-2, 2}b_{-2, 2}^*= \frac{g^2 (\text{n}_x-1) \text{n}_x \left(\text{n}_y^2+3 \text{n}_y+2\right) \hbar ^{10} \sin ^2\left(\frac{g t \hbar ^4}{4 M^4 \omega ^4}\right) \sin ^2\left(\frac{t \hbar  g (2 \text{n}_x-1) \hbar ^3 }{8 M^4 \omega ^4}+t \omega  \hbar\right)}{256 M^{10} \omega ^{10}} \nn \\
\nn\\
&&b_{0, 0}b_{0, 0}^*=\frac{\hbar ^4 }{4 M^2}  \cos ^2\left(\frac{g t \hbar ^4 (2 \text{n}_y +1)}{8 M^4 \omega ^4}+t \omega  \hbar \right)    \nn\\
\nn\\
&& b_{ 2,-2}b_{ 2,-2}^*  =\frac{g^2 \left(\text{n}_x +3 \text{n}_x +2\right) (\text{n}_y -1) \text{n}_y  \hbar ^{10} \sin ^2\left(\frac{g t \hbar ^4}{4 M^4 \omega ^4}\right) \sin ^2\left(\frac{g (2 \text{n}_x +3) t \hbar ^4}{8 M^4 \omega ^4}+t \omega  \hbar \right)}{256 M^{10} \omega ^{10}} \nn \\
\nn\\
&& b_{ 2, 2}b_{-2, 2}^* = \frac{g^2 \left(\text{n}_x ^2+3 \text{n}_x +2\right) \left(\text{n}_y ^2+3 \text{n}_y +2\right) \hbar ^{10} \sin ^2\left(\frac{g t \hbar ^4}{4 M^4 \omega ^4}\right) \left(5+3 \cos \left(\frac{g (2 \text{n}_x +3) t \hbar ^4}{4 M^4 \omega ^4}+2 t \omega  \hbar \right)\right)}{2048 M^{10} \omega ^{10}} \nn
\ee
while $b_{0,-2}b_{0,-2}^*$ and $b_{0,-2}b_{0,-2}^*$ are  given in appendix  B.  

The properties of OTOC formula  (\ref{Ccn}) are discussed in the following subsections.
%%%%%%%%%%%%%%%%%%%%%
%%%%%%%%%%%%%%%%%%%%%
\subsection{Microcanonical OTOC Properties of CHO}
We use  formula  (\ref{Ccn}) to plot   figure 1 to see the property of the microcanonical OTOC of coupled harmonic oscillators.   The  values of parameter we used are : $\omega= 0.5,\  \hbar =M=1,\ \text{n}_x=\text{n}_y=15, \ g=0.01$.  
\\
\\
\scalebox{0.6}{\hspace{1.5cm}\includegraphics{figure1}}
\\
  {{\bf Figure 1.} Microcanonical OTOC of coupled harmonic oscillators for $n_x=n_y=15$. The growth of OTOC shown in the early time  signatures the property associated to the quantum chaos.}  
%%%%%%%%%%%%%%%dd%%%%%%%%%%%%
\\

We make following comments to discuss the properties of OTOC shown in    figure 1 :

1. The zero order microcanonical OTOC is 
\be
c^{g=0}_{\vec n}(t)=\frac{\hbar ^4 \cos ^2(t \omega  \hbar )}{4 M^2}\to \frac{ \cos ^2(t /2)}{4 }\le 0.25   \label{0}
\ee
  However the  figure shows that OTOC will increase to  $c_{\vec n}(t)\approx$ 300 near at $t\approx40$.  At first sight the property   violates the conventional property of perturbation :  the quantity of the zero order  is larger then that of the leading order.     In fact, this is not the case in $c_{\vec n}(t)$. 

2. To explain this property let us first consider, for example, the term
\be
b_{0, 0}b_{0, 0}^*&=&\frac{\hbar ^4 }{4 M^2}  \cos ^2\left(\frac{g \cdot t \ \hbar ^4 (2 \text{n}_y +1)}{8 M^4 \omega ^4}+t \omega  \hbar \right)\approx  \frac{\hbar ^4 }{4 M^2}  \cos ^2\left(\frac{g \cdot t \ \hbar ^4 (2 \text{n}_y +1)}{8 M^4 \omega ^4} \right)
\ee
which is an approximation value if
\be
\frac{g \cdot   \ \hbar ^4 (2 \text{n}_y +1)}{8 M^4 \omega ^4}\gg   \omega  \hbar
\ee
This means that ``$g\cdot n_x$ " can become large for high level, i.e. $n_x\gg 1$, even ``$g$ " is small.

3.  Next,   consider  a correction term in appendix B :
\be
b_{0, 2}\ b_{0, 2}^*\sim   g \left(n_x+1\right) n_x \cos \left(\frac{g \cdot t \ \hbar ^4}{2 M^4 \omega ^4}\right)
\ee
There is a factor ``$   g\cdot t    $" inside cosine, which can become large for later time, i.e. $ g\cdot t \gg 1$ even ``$g$" is small. Thus  this term is an oscillation at later time. Consider the high level with $n_x\gg 1$ such that $g \left(n_x+1\right) n_x\gg 1$, then the correction term will be larger then zero order in (\ref{0}) therefore.

This factor plays a crucial role to  enhance the OTOC during time is evolving, even $g$ is small. The growth of OTOC shown in the early time  in the left-hand part of figure 1 signatures the property associated to the quantum chaos in coupled harmonic oscillators. 

4. The right-hand part of figure 1 shows a behavior of increasing  harmonic oscillators.  In fact, the  property of increasing oscillation could easily be found in a function with two oscillations with different frequency and different amplitude, for example  the function ``$ 1 + 24 \sin(  t)^2 + \cos(10\ t )$".
 
5.  The property that the  magnitude of enhancement  is proportional $n_x$  is clearly shown in Figure 2. Therefore, the enhancement   will not be shown when   $n_x$ is too small and the function of  microcanonical OTOC will be like as a simple oscillation, as shown in the left-hand side diagram in Figure 2. In this case the growth behavior of the quantum chaos will disappear therefore. Note that the property :  low modes behave periodic in time  while higher modes deviate from the periodic behavior, was first found   in numerical analysis in paper \cite{Hashimoto20b}.  Our study in perturbation can see this property too.
%%%%%%%%%%%%%%%%%%%%%%%%%
\\
\\
\scalebox{0.6}{\hspace{2cm}\includegraphics{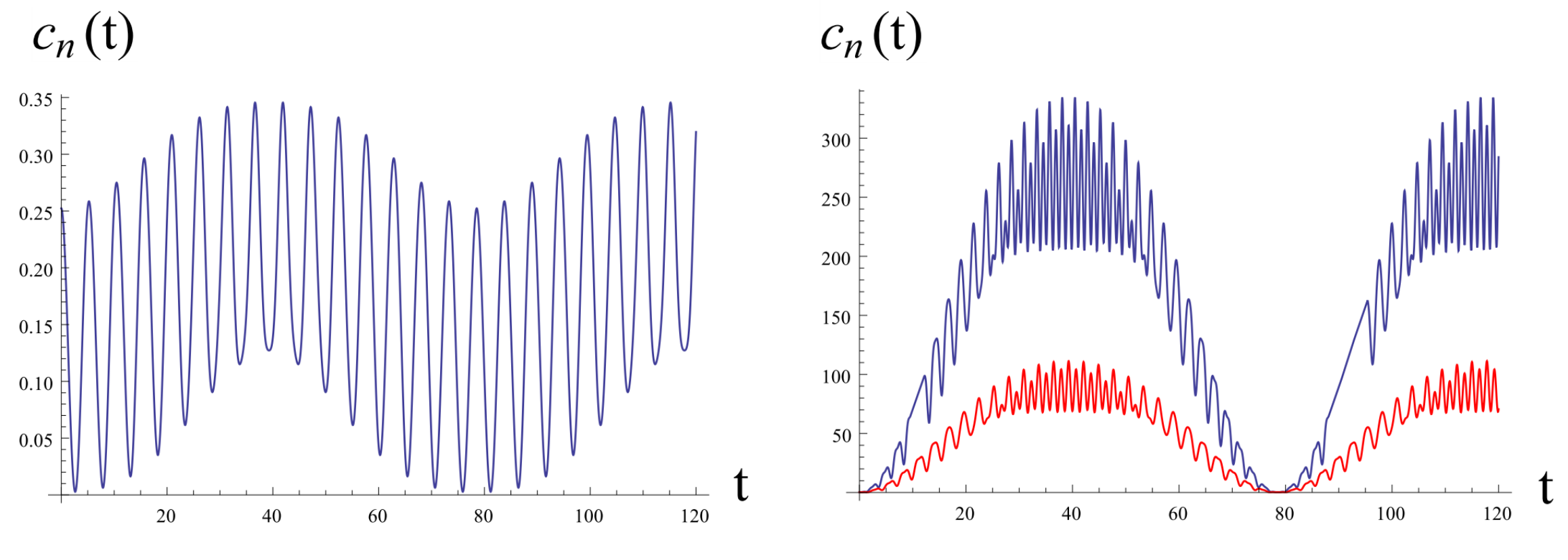}}
\\
  {{\bf Figure 2.} Microcanonical OTOC of coupled harmonic oscillators. Left-hand diagram : $n_x=n_y=2 $.  Left-hand diagram :  Blue (upper) describes CHO of  $n_x=n_y=20$, Red (lowerer) describes CHO of $n_x=n_y=15$. It shows that the  magnitude of enhancement  is proportional to $n_x,n_y$} 
%%%%%%%%%%%%%%%%%%%
\\
\\
6. Left-hand side diagram in Figure 2 shows that microcanonical OTOC, $c_{\vec n}$, is an increasing  oscillation function in initial.  Then, becomes a decreasing oscillation function. Then,  becomes an increasing oscillation function. And so on. Therefore, the property found  in numerical analysis \cite{Hashimoto20b} : ``at late time stage OTOC  saturates to a constant value"  does not show  in perturbation result in this paper. To see the property let us write the high level   relation  of $c_{\vec n}$ at  late time \footnote{We adopt the approximation, such as $\cos \left(\frac{g t \hbar ^4 (\text{n}_x+\text{n}_y+1)}{2 M^4 \omega ^4}+4 t \omega  \hbar \right)\approx \cos \left(\frac{g t \hbar ^4 (\text{n}_x+\text{n}_y )}{2 M^4 \omega ^4} \right)$, as $t,n_x,n_y\gg1$. }
\be
c_{\vec n} \approx \frac{g^2 \text{n}_x^2 \text{n}_y^2 \hbar ^{10} \sin ^2 (\frac{g t \hbar ^4}{4 M^4 \omega ^4} ) \left(18-\cos  (\frac{g  t \hbar ^4   \text{n}_x}{2 M^4 \omega ^4} )\right)}{1024 M^{10} \omega ^{10}},~~~t,n_x,n_y\gg1
\ee
Thus the high level  microcanonical OTOC  $c_{\vec n}$ is enhanced  at  late time which, however, is still a oscillation function without saturates to a constant value. 
\\
%%%%%%%%%%%%%%%%%%%
%%%%%%%%%%%%%%%%%%%
\subsection{Thermal OTOC Properties of CHO}
%%%%%%%%%%%%%%%%%%%
We can now substitute analytic form of  $c_n(t)$ in (\ref{Ccn}) into formula of $C_T(t)$ in (\ref{TC}) to numerically evaluate  the thermal OTOC of  coupled harmonic oscillators. The results are plotted in Figure 3 in which the  values of parameter we used are : $\omega= 0.5,\  \hbar =M=1, \ g=0.01$.  The summation can be done numerically up to  $\text{n}_x=\text{n}_y=100$ and is already accurate enough.  Note that  for higher temperature we need to sum more higher energy levels to have a accurate summation.
\\
\\
\scalebox{0.5}{\hspace{6cm}\includegraphics{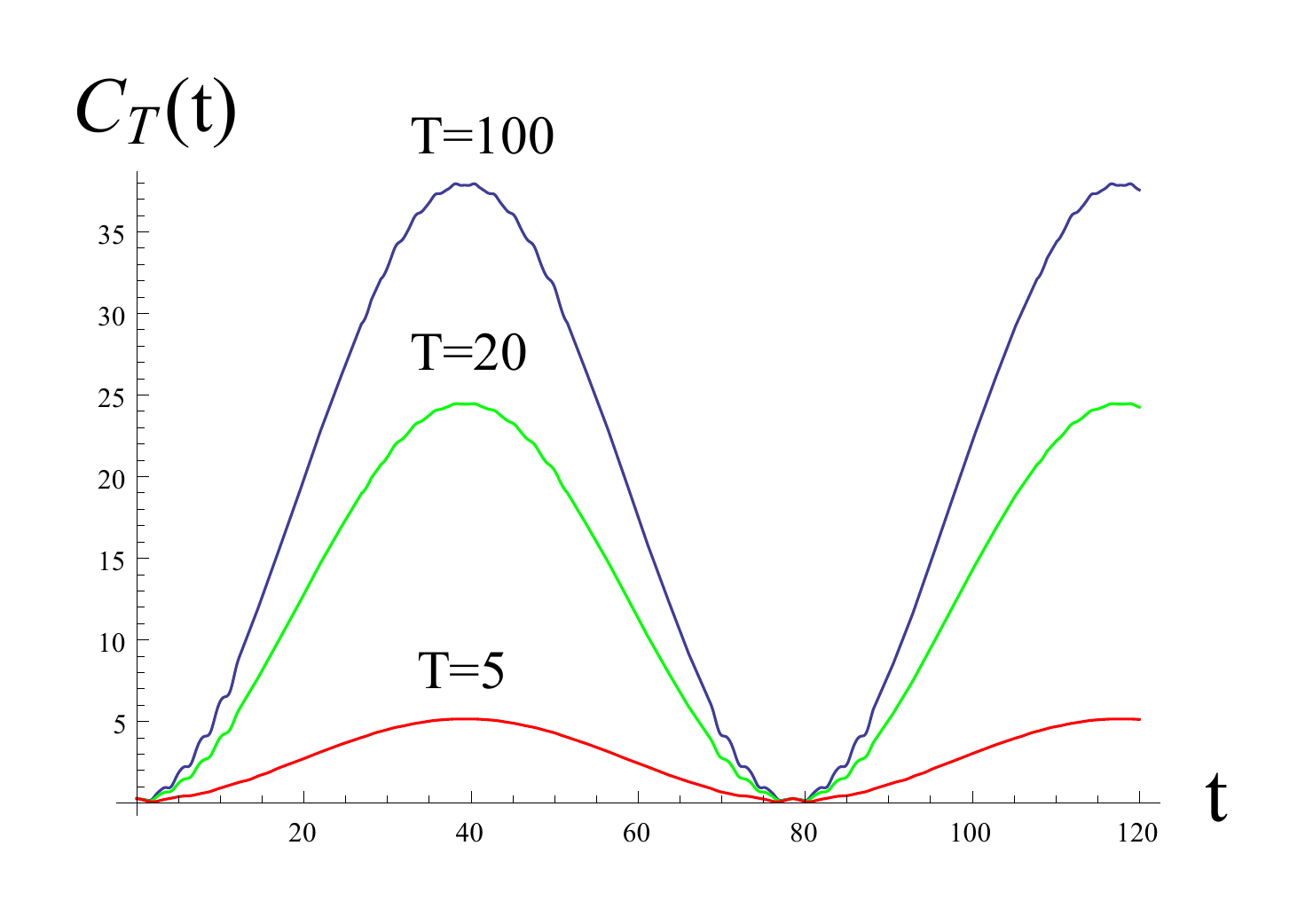}}
\\
  {{\bf Figure 3.} Thermal OTOC of coupled harmonic oscillators for T=100, 20, 5. It shows that the  magnitude of enhancement  is proportional to temperature $T$.}
\\

We make following comments to discuss the properties of thermal OTOC shown in    figure 3:

1.  The thermal OTOC is the thermal average of microcanonical OTOC in which we have to sum over all possible energy level, $n_x,n_y$.    As different level will oscillate with different frequency and sum over   all possible level will thus smooth the oscillation. In this way, the increasing oscillation behavior in $c_n(t)$, see figure 2, is now changed to a simple increasing behavior in figure 3.

2.  The figure 3  shows that the  magnitude of enhancement  is proportional to the temperature.  This is because that there is  the Boltzmann factor in calculating thermal OTOC.  The factor  tells us that the higher the temperature, the geater  the   energy level will contribute.  And, as discussed in figure 2, the higher level will have large  enhancement.  

3. Like that in  microcanonical OTOC  the thermal OTOC, $C_{T}(t)$, is an increasing   function in initial.  Then, becomes a decreasing   function. Then,  becomes an increasing   function. And so on. Therefore, the property found  in numerical analysis \cite{Hashimoto20b} : ``at late time stage OTOC  saturates to a constant value"  does not shown in perturbation result in this paper. 

4.  It is know that, after the Ehrenfest time, the value of the thermal OTOC is expected to approach asymptotically in time with relation  \cite{Maldacena16}
\be
C_T(\infty)=2\<x^2\>_T\<p^2\>_T
\ee 
This relation, which strongly indicates that the   Hamiltonian system is quantum chaotic, was shown in numerical analysis in \cite{Hashimoto20b}. However, our  perturbation OTOC  does not saturate  to a constant value. It fells short of author's expectation.

5. To clearly see the properties of thermal OTOC in initial state we plot Figure 4.
\\
\\
\scalebox{0.5}{\hspace{6cm}\includegraphics{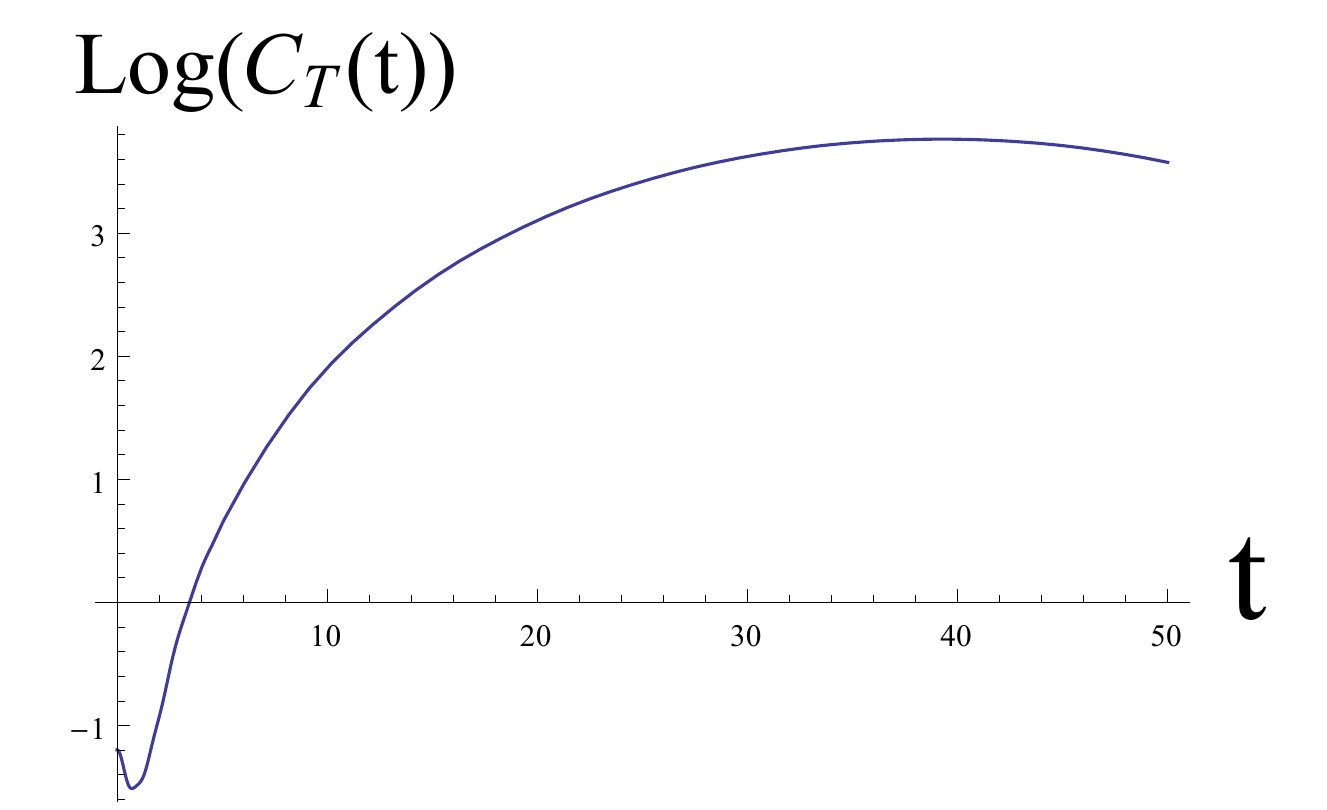}}
\\
 Figure 4 :  Log($C_T$(t))   as the function of time.  The growth of the OTOC being exponential is not found.
\\
\\
%%%%%%%%%%%%%%%%%%%
 which shows that the initial stage does not have a linear property of \text{Log($C_T$(t))}.  Therefore,  quantum chaos character of growth of the OTOC being exponential  does not show in the first-order perturbation studied in this paper, although  the enhance property discussed before does lead to a rapidly rising characteristic of the initial time of OTOC.  
%%%%%%%%%%%%%%%%%%%
%%%%%%%%%%%%%%%%%%%
%%%%%%%%%%%%%%%%%%%%%%%%%
%%%%%%%%%%%%%%%%%%%                   %%%%%%%%%%%%%%%%%%%%
%%%%%%%%%%%%%%%%%%%%                   %%%%%%%%%%%%%%%%%%%%
%%%%%%%%%%%%%%%%%%%%                   %%%%%%%%%%%%%%%%%%%%
%%%%%%%%%%%%%%%%%%%%                   %%%%%%%%%%%%%%%%%%%%
%%%%%%%%%%%%%%%%%%%%                   %%%%%%%%%%%%%%%%%%%%
%%%%%%%%%%%%%%%         Anharmonic Oscillator      %%%%%%%%%%%%%%%
%%%%%%%%%%%%%%%%%%%%                   %%%%%%%%%%%%%%%%%%%%
\section{Perturbative OTOC of Anharmonic Oscillator}
%%%%%%%%%%%%%%%%%%%%
Wavefunction approach to OTOC of anharmonic oscillator (AHO) had been studied in \cite{Romatschke} following the Hashimoto paper \cite{Hashimoto17}. In this section we will analyze the same topic  while  by the  second quantization approach in perturbative method.  This section is self-contained, and readers who are only interesting in AHO can skip section 3 and read this section directly.
%%%%%%%%%%%%%%%%%%%%%%%%%
\subsection{First-order Microcanonical OTOC of AHO}
%%%%%%%%%%%%%%%%%%
The anharmonic oscillators with quartic interaction is described by
\be
\tilde H&=&({p^2\over 2M}+{M\omega^2\over 2}x^2)+g{x^4}\nn\\
&=&  \hbar\omega \( a^\dag a+ \frac{1}{2}\) +{g } \ \({\hbar\over2M\omega}\)^2(a^\dag+a )^4=H +V
\ee
which has  a well-known unperturbed solution
\be
H |n \>=E_n |n \>=\hbar\omega\(n +{1\over2}\) |n  \> 
\ee
Use first-order perturbation formulas in quantum mechanics textbook 
\be
\widetilde {|n\>}&\approx& |n \>+\sum_{n\ne m}{|m \>\<m |  V    |n \> \over E_n -E_m } \>                     \\
\widetilde {E_n}& \approx& E_n +\<n |\, V  \,   |n  \> 
\ee
we can calculate the leading corrections to state and energy. A basic quantity we need, after calculation, is
\be
V|n\>&=& \frac{g \hbar ^4 \sqrt{n-3} \sqrt{n-2} \sqrt{n-1} \sqrt{n} }{16 M^4 \omega ^4}\  |n -4 \>+\frac{g \hbar ^4 \sqrt{n-1} \sqrt{n} (2 n-1)   }{8 M^4 \omega ^4}\ |n -2 \>  \nn\\
\nn\\
&&  +\frac{3 g \hbar ^4 (2 n (n+1)+1)   }{16 M^4 \omega ^4}\  |n   \>+\frac{g \hbar ^4 \sqrt{n+1} \sqrt{n+2} (2 n+3)   }{8 M^4 \omega ^4}\    |n +2 \>                      \\
\nn\\
&& +\frac{g \hbar ^4 \sqrt{n+1} \sqrt{n+2} \sqrt{n+3} \sqrt{n+4}   }{16 M^4 \omega ^4}                         \    |n +4 \> 
\ee
% %%%%%%%%%%%%%%%%%%%%
The first-order energy $\widetilde {E_n}$ and Fock state $\widetilde {|n\>}$ become
\be
 \widetilde {E_n} &\approx&  E_n +\<n |\, {g } \ \({\hbar\over2M\omega}\)^2(a^\dag+a )^4  \,   |n  \>=\hbar \omega \(n+{1\over2}\) +\frac{3 g \hbar ^4 (2 n (n+1)+1) }{16 M^4 \omega ^4} ~~\label{E1}  \\
\nn\\
\widetilde {|n\>}&\approx&   |n  \> +\frac{g  \hbar ^3 \sqrt{n-3} \sqrt{n-2} \sqrt{n-1} \sqrt{n}  }{64 M^4 \omega ^5}\ |n -4 \>+\frac{g  \hbar ^3 \sqrt{n-1} \sqrt{n} (2 n-1)  }{16 M^4 \omega ^5}\ |n -2 \>  \nn\\
\nn\\
&&-\frac{g  \hbar ^3 \sqrt{n+1} \sqrt{n+2} (2 n+3)  }{16 M^4 \omega ^5}\ |n +2 \>-\frac{g  \hbar ^3 \sqrt{n+1} \sqrt{n+2} \sqrt{n+3} \sqrt{n+4}  }{64 M^4 \omega ^5}\ |n +4 \> ~~~~~
\ee
%%%%%%%%%%%%%%%%%%%%%%%%%
%%%%%%%%%%%%%%%%%%
To proceed, we have  an interesting and simple relation
\be
x  \widetilde{| n\>}&=&g  \xi_{-3}(n)  \widetilde{| n-3\>}+(  \kappa_{-1}(n)+g  \xi_{-1}(n))   \widetilde{| n-1\>}+(  \kappa_{1 }(n)+g  \xi_{1}(n))   \widetilde{| n+1\>}\nn\\
&&+g  \xi_{3}(n)   \widetilde{| n+3\>}+{\cal O}(g^2) 
\ee
where
\be
\kappa_{-1}(n)&=&   \sqrt{\hbar\over2M\omega}\ \sqrt{n},~~~~~~~~~~~~~~~~~~~~~~~~~\kappa_{1}(n)=      \sqrt{\hbar\over2M\omega}\ \sqrt{n+1}        \\
\xi_{-1}(n)&=&-\frac{  \hbar ^3 \sqrt{\frac{\hbar }{2M \omega }} \ n^{3/2}  }{8  M^4 \omega ^5},~~~~~~~~~~~~~~~~~~~~          
\xi_{ 1}(n) =   -\frac{  \hbar ^3  \sqrt{\frac{\hbar }{2M \omega }}\  (n+1)^{3/2}\    }{8  M^4 \omega ^5}                  \\
\xi_{-3}(n)&=&\frac{  \hbar ^3   \sqrt{\frac{\hbar }{2M \omega }}\ \sqrt{n-2} \sqrt{n-1} \sqrt{n}\      }{16   M^4 \omega ^5},  ~~~    \xi_{3}(n) =   \frac{  \hbar ^3  \sqrt{\frac{\hbar }{2M \omega }} \ \sqrt{n+1} \sqrt{n+2} \sqrt{n+3} }{16   M^4 \omega ^5}  ~~~~               
\ee
Therefore
\be
\widetilde{x}_{nm}= g  \xi _{-3}(m) \delta _{n,m-3}+\left(\kappa _{-1}(m)+g  \xi _{-1}(m)\right) \delta _{n,m-1}+\left(\kappa _1(m)+g  \xi _1(m)\right) \delta _{n,m+1}+g  \xi _3(m) \delta _{n,m+3}~~~\label{Axnm}\nn\\
\ee
Substituting spectrum   $\widetilde{E}_n$ in (\ref{E1}) and  matrix element  $\widetilde{x}_{nm}$ in (\ref{Axnm}) into formula $b_{nm}$ in  (\ref{bnm})  we obtain
\be 
 \widetilde{b}_{nm}&=& {1\over 2}\sum_k\ \widetilde{x}_{nk}\widetilde{x}_{km}\(e^{i\widetilde{E} _{nk}t} \widetilde{E}_{km} -e^{i\widetilde{E}_{km}t}\widetilde{E}_{nk}\)  \nn \\
&=&\widetilde{B}_{n,m+4} \ \delta_{n,m+4}+\widetilde{B}_{n,m+2} \ \delta_{n,m+2}+\widetilde{B}_{n,m } \ \delta_{n,m }+\widetilde{B}_{n,m-2} \ \delta_{n,m-4}+\widetilde{B}_{n,m-4} \ \delta_{n,m-4}\nn  \label{bA}\\
\ee
where $  \widetilde{B}_{i,j}$ are given in appendix C.  It is seen that that $\widetilde{b}_{i,j}$ dominates by nearest-neighbor and next  nearest-neighbor energy levels. However,  the next  nearest-neighbor energy levels, terms of  $ \widetilde{B}_{n,m+4}$ and $ \widetilde{B}_{n,m-4}$, are proportional to $g$ and, thus, their contributions to  the microcanonical OTOC $c_n(t)$ is order of $g^2$. Therefore, in first order approximation    they  shall be dropped off   and we can conclude that  microcanonical OTOC dominates by nearest-neighbor.  This property is mentioned in \cite{Romatschke} in which  the assumption of small value of potential is not made.  
\\
 
To leading order of “$g$" only the nearest-neighbor energy level contributes and explicitly values of  $\widetilde{b}_{nm}$ leads to 
\be
\widetilde{c}_n(t)&=&\sum_m(i\widetilde{b}_{nm})(i\widetilde{b}_{nm})^* =  |\widetilde{B}_{n,n-2}|+|\widetilde{B}_{n,n}|+|\widetilde{B}_{n,n+2}\nn| \\
&=& \frac{1  }{64 M^6 \omega ^5}(s0 + s1 + s2 + s3 + s4 + s5 + s6 + s7 + s8)  ~~~~ ~~~~\label{cn1}
\ee
where  
\be
s0&=&   3 g \left(4 \left(n^3+n\right)+1\right) \hbar ^7+2 M^4 (4 n (n+1)+3) \omega ^5 \hbar ^4      \nn             \\
s1&=&  -2 \hbar ^4 \left(3 g \left(2 n^3+n\right) \hbar ^3+2 M^4 (2 n (n+1)+1) \omega ^5\right) \cos \left(\frac{3 g \hbar ^4  t \ }{4 M^4 \omega ^4}\right)          \nn              \\
s2&=&    \frac{1}{4} n \hbar ^4 \left(g \left(11 n^2+1\right) \hbar ^3+8 M^4 n \omega ^5\right) \cos \left( \frac{3 g \hbar ^4 t  n \ }{2 M^4 \omega ^4}+2  \hbar \omega t \right)        \nn              \\
s3&=&   -\frac{1}{4} n (n+1) \hbar ^4 \left(11 g (2 n+1) \hbar ^3+16 M^4 \omega ^5\right) \cos \left(\frac{3 g \hbar ^4  (2 n+1) t  }{4 M^4 \omega ^4}+2  \hbar \omega t  \right)         \nn              \\
s4&=&     -\frac{1}{4} g (n-2) (n-1) n \hbar ^7 \cos \left(\frac{3g \hbar ^4  (n-1) t  }{2 M^4 \omega ^4}+2 \hbar \omega t  \right)             \nn        \\
s5&=&  \frac{1}{4} (n+1) \hbar ^4 \left(g (11 n (n+2)+12) \hbar ^3+8 M^4 (n+1) \omega ^5\right) \cos \left(\frac{3 g \hbar ^4  (n+1) t  }{2 M^4 \omega ^4}+2  \hbar \omega t \right)       \nn               \\
s6&=&         -\frac{1}{4} g (n+1) (n+2) (n+3) \hbar ^7 \cos \left(\frac{3 g \hbar ^4  (n+2) t  }{2 M^4 \omega ^4}+2  \hbar \omega t  \right)         \nn        \\
s7&=&     \frac{1}{4} g (n-2) (n-1) n \hbar ^7 \cos \left(\frac{3g \hbar ^4  (2 n-3) t  }{4 M^4 \omega ^4}+2  \hbar \omega t  \right)            \nn         \\
s8&=&  \frac{1}{4} g (n+1) (n+2) (n+3) \hbar ^7 \cos \left(\frac{3g \hbar ^4  (2 n+5) t }{4 M^4 \omega ^4}+2   \hbar \omega t \right)         \nn               
\ee
in which $s1\sim s8$ are harmonic modes with different frequencies.  
%%%%%%%%%%%%%%%%%%%%%
\subsection{OTOC Properties of AHO}
%%%%%%%%%%%%%%%%%%
We plot   figure 5 to see how the  microcanonical OTOCs,   $\widetilde{c}_n(t)$, evolutes with time in below : 
\\
\\
\scalebox{0.6}{\hspace{1.5cm}\includegraphics{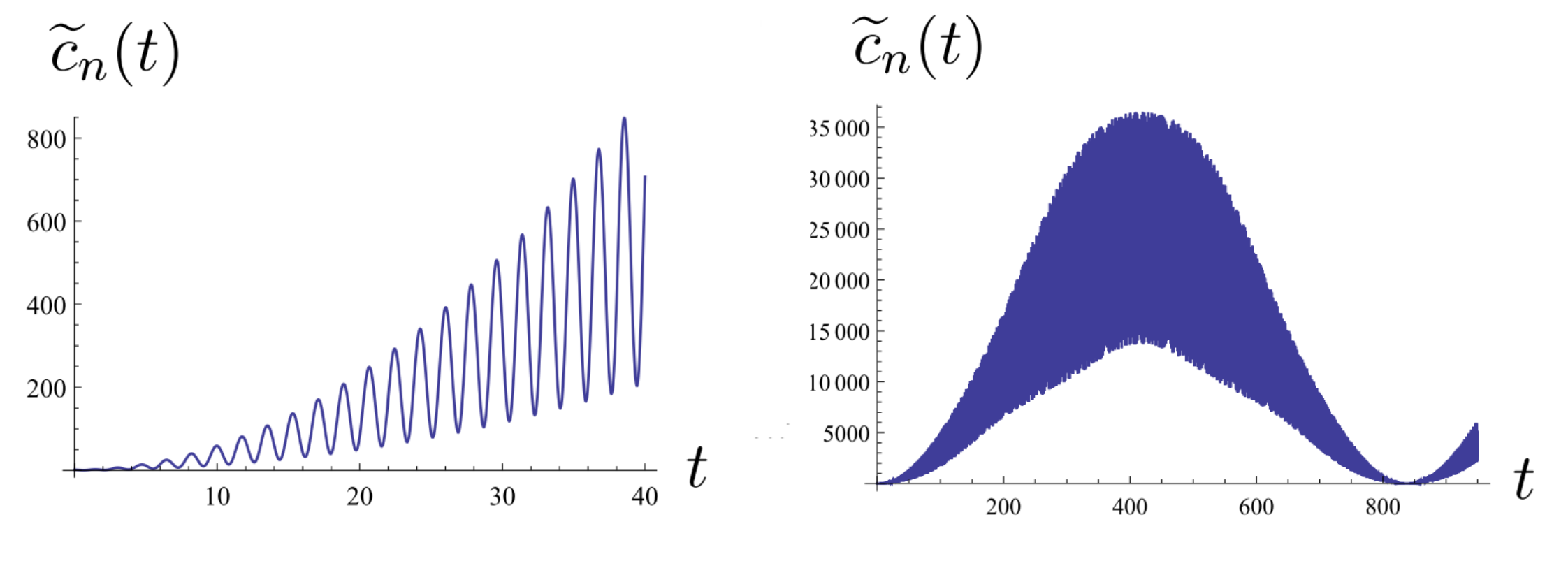}}
\\
{Figure 5:  Microcanonical OTOCs  $\widetilde{c}_n(t)$ described in (\ref{cn1})  as the function of time. The parameters are chosen by $\omega =\hbar=M=1$ while $g=0.01$ and $n=100$.}
\\

The figure 5 shows the  property likes as that in CHO : Microcanonical OTOCs is an increasing oscillation function in initial.  Then, becomes a decreasing oscillation function. Then,  becomes an increasing oscillation function. And so on.   To see the property of  OTOC $\widetilde{c}_n(t)$ at late time one can  use  a relation 
\be
\widetilde{c}_n(t)\approx\frac{3 g n^3 \hbar ^7 \sin ^2\left(\frac{3 g t \hbar ^4}{8 M^4 \omega ^4}\right)}{8 M^6 \omega ^5},~~~t,n \gg1
\ee
Thus, the high level  microcanonical OTOC  $\widetilde{c}_n(t)$  is enhanced  at  late time which, however, is still a oscillation function without saturating to a constant value, as that in CHO. In this way, the thermal OTOC will have same behaviors   as those in CHO.
%%%%%%%%%%%%%%%%%%%
 \section{Conclusions}
In this paper we use the method which  was set up  by Hashimoto recently in \cite{Hashimoto17,Hashimoto20a,Hashimoto20b} to calculate the OTOC in the  systems of non-linearly coupled oscillators (CHO) and  anharmonic (quartic) oscillators (AHO). In contrast to the previous studies we  calculate  the OTOC  by second quantization method in  perturbative approximation  , which can gives  analytical formulas. 

Some interesting properties could be read from the  perturbation results. For example,  using the analytical formulas we clearly see  that the interactions can enhance the correlation to very large values over time and is proportional to the energy level.  The rapidly rising characteristic of the initial time of OTOC is also clearly shown. While this signatures the  quantum chaos  the exponential growth of the OTOC that indicates
quantum chaoticity does not found in   the first-order perturbation studied in this paper.  The microcanonical OTOC which is an  increasing oscillation  becomes a simple increasing function in thermal OTOC, as the thermal OTOC  is a summation over all energy level of   microcanonical OTOC and  different level   oscillates with different frequency.  

The enhancement property shown in the analytic formula tells us that the appearing quantity ``$g\cdot t$'' and ``$g\cdot n$'' could render OTOC to be very large  at later time and/or in large level even the coupling ``g'' is very small  This property reveals that, more or less, the small coupling value of ``g''  could capture the physical characters of general strength of coupling ``g''.  Therefore,  while the property that, at late time stage OTOC  saturates to a constant value does not show in the first-order perturbation studied in this paper  one could  expect that the higher-order perturbation will be able to  find  the saturation property. 

Finally,  the problem of many-body chaos at weak coupling had been investigated in several years ago by Stanford \cite{Stanford}, in which the system of matrix $\Phi^4$ theory was studied. In a recent paper, Kolganov and Trunin  detailly studied the  classical and quantum butterfly effect in a related theory \cite{Kolganov}.  Since that the interacting scalar field theory could be transformed to a system of coupled oscillators, see for example \cite{Huang21}, it is interesting to study the problems along the prescription of this paper.
%%%%%%%%%%%%%%%%%%%
\newpage
%%%%%%%%%%%%%%%%%%%  
\appendix
\section{$b_{0,-2}$ and $b_{0, 2} $}
In this appendix, we  list  the coefficients of $b_{0,-2}$ and $b_{0, 2} $  in (\ref{Cbxx}): 
\be
b_{0,-2}&=&-\frac{1}{64 M^5 \omega ^5}g \hbar ^5 \exp \left(\frac{i g t \hbar ^4 \left(2 n_x+1\right)}{8 M^4 \omega ^4}+i t \omega  \hbar \right)\left[2 n_x \sqrt{n_y+1} \sqrt{n_y+2} \cos \left(\frac{g t \hbar ^4 \left(n_x+n_y-1\right)}{4 M^4 \omega ^4}+2 t \omega  \hbar \right)\right.\nn
\\
&&-2 \left(n_x+1\right) \sqrt{n_y+1} \sqrt{n_y+2} \cos \left(\frac{g t \hbar ^4 \left(n_x+n_y+1\right)}{4 M^4 \omega ^4}+2 t \omega  \hbar \right)\nn\\
&&-i \left(2 n_x \sqrt{n_y-1} \sqrt{n_y} \sin \left(\frac{g t \hbar ^4 \left(n_x-n_y\right)}{4 M^4 \omega ^4}\right)-2 \left(n_x+1\right) \sqrt{n_y-1} \sqrt{n_y} \sin \left(\frac{g t \hbar ^4 \left(n_x-n_y+2\right)}{4 M^4 \omega ^4}\right)\right.\nn\\
&&+\left.\left.\sqrt{n_y+1} \sqrt{n_y+2} \left(n_x \sin \left(\frac{g t \hbar ^4 \left(n_x+n_y-1\right)}{4 M^4 \omega ^4}+2 t \omega  \hbar \right)-\left(n_x+1\right) \sin \left(\frac{g t \hbar ^4 \left(n_x+n_y+1\right)}{4 M^4 \omega ^4}+2 t \omega  \hbar \right)\right)\right)\right]\nn\\
\nn\\
\nn\\
b_{0, 2}&=&\frac{1}{128 M^5 \omega ^5}g  \left[2 n_x \exp \left(-     \frac{ig t \hbar ^4 \left(2 n_y+5\right)}{M^4}-i t \omega  \hbar \right) \left(-1+e^{\frac{i g t \hbar ^4 \left(-n_x+n_y+2\right)}{2 M^4 \omega ^4}}\right)\right.\nn\\
&&  - n_x \left(-\exp \left(-\frac{i g t \hbar ^4 \left(4 n_x+2 n_y+3\right)}{8 M^4 \omega ^4}-3 i t \omega  \hbar \right)\right) \left(1+3 \exp \left(\frac{i g t \hbar ^4 \left(n_x+n_y+1\right)}{2 M^4 \omega ^4}+4 i t \omega  \hbar \right)\right)                  \nn\\
&&-2 \left(n_x+1\right) \exp \left(-    \frac{ig t \hbar ^4 \left(2 n_y+1\right)}{M^4} -i t \omega  \hbar \right) \left(-1+e^{-\frac{i g t \hbar ^4 \left(n_x-n_y\right)}{2 M^4 \omega ^4}}\right)                     \nn\\
&&    +\left.\left(n_x+1\right) +\exp \left(\frac{i g t \hbar ^4 \left(2 n_y+5\right)}{8 M^4 \omega ^4}+i t \omega  \hbar \right) \left(3+\exp \left(-    \frac{ig t \hbar ^4 \left(n_x+n_y+3\right)}{M^4}-i 4 t \omega  \hbar \right)\right)\right]                
\ee
%%%%%%%%%%%%%%%%%%%%%%%%%%%%
\section{$b_{0,-2}b_{0,-2}^*$ and $b_{0,-2}b_{0,-2}^*$}
In this appendix, we  list  the coefficients of $b_{0,-2}b_{0,-2}^*$ and $b_{0,-2}b_{0,-2}^*$ in (\ref{Ccn}): 
\be
&&b_{0,-2}b_{0,-2}^*=-\frac{g^2 \hbar ^{10}}{8192 M^{10} \omega ^{10}}\[ -\left(2 n_x^2+2 n_x+1\right) \left(9 n_y^2+11 n_y+10\right) +4 n_x^2 \left(n_y-1\right) n_y \cos \left(\frac{g t \hbar ^4 \left(n_x-n_y\right)}{2 M^4 \omega ^4}\right)\nn \\
&&+ 4 \left(n_x+1\right){}^2 \left(n_y-1\right) n_y \cos \left(\frac{g t \hbar ^4 \left(n_x-n_y+2\right)}{2 M^4 \omega ^4}\right)-8 n_x \left(n_x+1\right) \left(n_y-1\right) n_y \cos \left(\frac{g t \hbar ^4 \left(n_x-n_y+1\right)}{2 M^4 \omega ^4}\right)             \nn   \\
&&+4 n_x^2 \sqrt{\left(n_y-1\right) n_y \left(n_y+1\right) \left(n_y+2\right)} \cos \left(\frac{    g t \hbar ^4 \left(2 n_x-1\right) }{4 M^4 \omega ^4}+2 t \omega  \hbar\right) \nn\\
&&-3 n_x^2 \left(n_y^2+3 n_y+2\right) \cos  \(\frac{g t \hbar ^4 \left(n_x+n_y-1\right)}{2 M^4 \omega ^4}+4 t \omega  \hbar  \)                +6 n_x \left(n_x+1\right) \left(n_y^2+3 n_y+2\right) \cos \(\frac{  g t \hbar ^4 \hbar ^3 \left(n_x+n_y\right) }{2 M^4 \omega ^4} +4 t \omega  \hbar\)\nn\\
&&-3 \left(n_x+1\right){}^2 \left(n_y^2+3 n_y+2\right) \cos \left(\frac{g t \hbar ^4 \left(n_x+n_y+1\right)}{2 M^4 \omega ^4}+4 t \omega  \hbar \right)      +2 n_x \left(n_x+1\right) \left(9 n_y^2+11 n_y+10\right) \cos \left(\frac{g t \hbar ^4}{2 M^4 \omega ^4}\right) \nn\\
&&-8 n_x \left(n_x+1\right) \sqrt{n_y \left(n_y^3+2 n_y^2-n_y-2\right)} \cos \left(\frac{    g t \hbar ^4 \left(2 n_x+1\right) }{4 M^4 \omega ^4}+2 t \omega  \hbar\right)            \nn   \\
&&  +4 n_x \left(n_x+1\right) \sqrt{n_y \left(n_y^3+2 n_y^2-n_y-2\right)} \cos \left(\frac{   g t \hbar ^4 \left(2 n_y-3\right) }{4 M^4 \omega ^4}+2 t \omega  \hbar\right)\nn\\
&&+4 n_x \left(n_x+1\right) \sqrt{n_y \left(n_y^3+2 n_y^2-n_y-2\right)} \cos \left(\frac{    g t \hbar ^4 \left(2 n_y+1\right)  }{4 M^4 \omega ^4}+2 t \omega  \hbar\right)               \nn  \\
&& +4 \left(n_x+1\right){}^2 \sqrt{n_y \left(n_y^3+2 n_y^2-n_y-2\right)} \cos \left(\frac{   g t \hbar ^4 \left(2 n_x+3\right) }{4 M^4 \omega ^4}+2 t \omega  \hbar\right)\nn\\
&&-4 \left(2 n_x^2+2 n_x+1\right) \sqrt{n_y \left(n_y^3+2 n_y^2-n_y-2\right)} \cos\(\frac{   g t \hbar ^4 \left(2 n_y-1\right) }{4 M^4 \omega ^4}+2 t \omega  \hbar\) \] 
 \ee
and
\be
&&b_{0, 2}b_{0, 2}^*= -\frac{g^2 \left(\text{n}_y ^2+3 \text{n}_y +2\right) \hbar ^{10}}{8192 M^{10} \omega ^{10}}\[ -9 \left(2 \text{n}_x ^2+2 \text{n}_x +1\right)  +4 n_x^2 \cos \left(\frac{    g t \hbar ^4 \left(2 n_x-1\right) }{4 M^4 \omega ^4}+2 t \omega  \hbar\right)\nn\\
&& +18 \left(n_x+1\right) n_x \cos \left(\frac{g t \hbar ^4}{2 M^4 \omega ^4}\right)+ 4 n_x^2 \cos \left(\frac{g t \hbar ^4 \left(-n_x+n_y+2\right)}{2 M^4 \omega ^4}\right)-3 n_x^2 \cos \left(\frac{g t \hbar ^4 \left(n_x+n_y+1\right)}{2 M^4 \omega ^4}+4 t \omega  \hbar \right)             \nn   \\
&&-8 n_x \left(n_x+1\right) \cos \left(\frac{    g t \hbar ^4 \left(2 n_x+1\right)  }{4 M^4 \omega ^4}+2 t \omega  \hbar\right)-8 n_x \left(n_x+1\right) \cos \left(\frac{g t \hbar ^4 \left(n_x-n_y-1\right)}{2 M^4 \omega ^4}\right)               \nn  \\
&&  +6 n_x \left(n_x+1\right) \cos \left(\frac{  \left(g t \hbar ^4 \left(n_x+n_y+2\right) \right)}{2 M^4 \omega ^4}+2 t \omega  \hbar\right)+4 n_x \left(n_x+1\right) \cos \left(\frac{    g t \hbar ^4 \left(2 n_y+1\right)  }{4 M^4 \omega ^4}+2 t \omega  \hbar\right)             \nn  \\
&&  + 4 \left(n_x+1\right){}^2 \cos \left(\frac{    g t \hbar ^4 \left(2 n_x+3\right) }{4 M^4 \omega ^4}+2 t \omega  \hbar\right)+ 4 n_x \left(n_x+1\right) \cos \left(\frac{    g t \hbar ^4 \left(2 n_y+5\right)  }{4 M^4 \omega ^4}+2 t \omega  \hbar\right)           \nn   \\
&& + 4 \left(n_x+1\right){}^2 \cos \left(\frac{g t \hbar ^4 \left(n_x-n_y\right)}{2 M^4 \omega ^4}\right)-3 \left(n_x+1\right){}^2 \cos \left(     \frac{g t \hbar ^4 \left(n_x+n_y+3\right)}{2 M^4 \omega ^4}   +4 t \omega  \hbar\right)\nn\\
&&-4 \left(2 n_x^2+2 n_x+1\right) \cos \(\frac{    g t \hbar ^4 \left(2 n_y+3\right)  }{4 M^4 \omega ^4} +2 t \omega  \hbar\)\]  
\ee
%%%%%%%%%%%%%%%%%
\section{$\tilde B_{n,m}$}
In this appendix, we  list  the coefficients of $ \widetilde{B}_{i,j }$  in (\ref{bA}) : 
\be
 \widetilde{B}_{n,m }&=&  \frac{\hbar^2}{16 M^5 \omega ^5}\ e^{-\frac{3 i g\hbar ^4 (2 n+1) t }{4 M^4 \omega ^4}-i   \hbar  \omega t}  \(  -n \left(3 g   \hbar ^3  n+4 M^4 \omega ^5\right)  e^{\frac{3 i g (n+1) t \hbar ^4}{4 M^4 \omega ^4}}   (1+e^{\frac{6 i g \hbar ^4  n t}{4 M^4 \omega ^4}+2 i \hbar \omega  t  })    \nn                \\
&& +(n+1) \left(3 g \hbar ^3 (n+1)+4 M^4 \omega ^5\right) e^{\frac{3 i g n t \hbar ^4}{4 M^4 \omega ^4}}(1+e^{\frac{3 i g (2 n+2) t \hbar ^4}{4 M^4 \omega ^4}+2 i \hbar \omega t}\)         \ \delta_{n,m }\\
 \widetilde{B}_{n,m+4}&=& g \hbar ^5 \sqrt{n-3} \sqrt{n-2} \sqrt{n-1} \sqrt{n} \  \frac{1}{64 M^5 \omega ^5}\ e^{\frac{3 i g (n-6) \hbar ^4  t}{4 M^4 \omega ^4}+i  \hbar \omega t }\\
&&\times \(-1+e^{\frac{9 i g  \hbar ^4  t}{4 M^4 \omega ^4}}\)\(3 \ e^{\frac{9 i g  \hbar ^4 t}{4 M^4 \omega ^4}}+e^{\frac{3 i g n  \hbar ^4  t}{2 M^4 \omega ^4}+2 i  \hbar \omega t } \)\ \delta_{n,m+4}\\
 \widetilde{B}_{n,m-4}&=&  g \hbar ^5 \sqrt{n-3} \sqrt{n-2} \sqrt{n-1} \sqrt{n} \ \frac{1}{64 M^5 \omega ^5}\ e^{-\frac{9 i g \hbar ^4 (n+3)  t}{4 M^4 \omega ^4}-3 i  \hbar \omega t }\\
&&\times \(-1+e^{\frac{9 i g   \hbar ^4 t}{4 M^4 \omega ^4}}\)\(1+3\ e^{ \frac{3 i g \hbar ^4 (2 n+5)  t}{4 M^4 \omega ^4}+2 i\hbar \omega   t }\)\ \delta_{n,m-4}\\
\tilde B_{n,m+2}&=&\frac{1}{64 M^5 \omega ^5}\sqrt{n-1} \sqrt{n}\ e^{  (-\frac{3 i g (n+1) t \hbar ^4}{4 M^4 \omega ^4}-i t \omega  \hbar t)}\nn\\
&& \hbar ^2\(-3 e^{\frac{9 i g t \hbar ^4}{4 M^4 \omega ^4}}g (n-2) \hbar ^3-e^{ (\frac{3 i g (2 n-1) t \hbar ^4}{2 M^4 \omega ^4}+4 i t \omega  \hbar )}g (n-2) \hbar ^3+3 g (n+1) \hbar ^3 \nn \\
&&+ e^{ (\frac{3 i g (4 n+1) t \hbar ^4}{4 M^4 \omega ^4}+4 i t \omega  \hbar ) } g (n+1) \hbar ^3  +2e^{(\frac{3 i g (2 n+1) t \hbar ^4}{4 M^4 \omega ^4}+2 i t \omega  \hbar)}(3 g (4 n-3) \hbar ^3+8 M^4 \omega ^5)\nn\\
&&-2e^{\frac{3 i g n t \hbar ^4}{2 M^4 \omega ^4}+2 i t \omega  \hbar } (3 g (4 n-1) \hbar ^3+8 M^4 \omega ^5)\)\\
\tilde B_{n,m-2}&=&-\frac{1}{64 M^5 \omega ^5} \sqrt{n+1} \sqrt{n+2}\ e^{(-\frac{i t \hbar  \left(3 g (4 n+7) \hbar ^3+16 M^4 \omega ^5\right)}{4 M^4 \omega ^4})}  \nn  \\
&& \hbar^2\(3g n \hbar ^3 e^{(\frac{3 i g (5 n+7) t \hbar ^4}{4 M^4 \omega ^4}+5 i t \omega  \hbar )}+g n \hbar ^3e^{\frac{3 i g (n+4) t \hbar ^4}{4 M^4 \omega ^4}+i t \omega  \hbar }\nn   \\
&& -   g (n+3) \hbar ^3 e^{\frac{3 i g (n+1) t \hbar ^4}{4 M^4 \omega ^4}+i t \omega  \hbar }+3 g (n+3) \hbar ^3 e^{\frac{15 i g (n+2) t \hbar ^4}{4 M^4 \omega ^4}+5 i t \omega  \hbar }      \nn             \\
&&+2  ( (3 g \hbar ^3+8 M^4 \omega ^5 ) e^{\frac{3 i g t \hbar ^4}{4 M^4 \omega ^4}}+3 g \hbar ^3-8 M^4 \omega ^5 ) e^{(\frac{3 i g (3 n+5) t \hbar ^4}{4 M^4 \omega ^4}+3 i t \omega  \hbar }\)
\ee
%%%%%%%%%%%%%%%%%%%%%%%%%%%
%%%%%%%%%%%%%%%%%%%
\\
\begin{center} 
{\bf  \large References}
\end{center}
%%%%%%%%%%%%%%%%%%%%%%
 
%%%%%%%%%%%%%%%%%%%
\end{document}